\documentclass[acmsmall]{acmart}
\pdfoutput=1

\usepackage{paralist}
\usepackage{booktabs}
\usepackage{color}

\newcommand{\dist}{\ensuremath{\text{dist}}}

\newcommand{\vset}{vertexSubset}

\newcommand{\emap}{\textproc{edgeMap}}
\newcommand{\emapdata}{\textproc{edgeMapData}}

\newcommand{\vmap}{\textproc{vertexMap}}

\newcommand{\emapsparse}{\textproc{edgeMapSparse}}
\newcommand{\emapblock}{\textproc{edgeMapBlocked}}

\newcommand{\cas}{CAS}
\newcommand{\ts}{test\_and\_set}
\newcommand{\fa}{fetch\_and\_add}
\newcommand{\writemin}{writeMin}

\newcommand{\makebkt}{\textproc{makeBuckets}}
\newcommand{\nextbkt}{\textproc{nextBucket}}

\newcommand{\updatebkt}{\textproc{updateBuckets}}

\newcommand{\orderinc}{\textsc{increasing}}

\newcommand{\nullbkt}{\textsc{nullbkt}}

\newcommand{\codevar}[1]{\mathit{#1}}

\newcommand{\src}{\textsc{src}}
\newcommand{\NC}{\mathsf{NC}}

\newcommand{\Boruvka}{Bor\r{u}vka}

\newcommand{\tsmod}{$\mathsf{TS}$}
\newcommand{\pwmod}{$\mathsf{PW}$}
\newcommand{\famod}{$\mathsf{FA}$}
\newcommand{\ram}{$\mathsf{RAM}$}
\newcommand{\mpram}{$\mathsf{TRAM}$}
\newcommand{\pram}{$\mathsf{PRAM}$}
\newcommand{\crcwpram}{$\mathsf{CRCW}\ \mathsf{PRAM}$}

\newcommand{\process}{thread}
\newcommand{\processes}{threads}

\DeclareMathAlphabet{\mathbfsf}{\encodingdefault}{\sfdefault}{bx}{n}

\usepackage{amsmath,algorithmicx,amssymb,subfigure,graphicx,url,multirow}
\usepackage{algorithm}
\usepackage[noend]{algpseudocode}
\usepackage{enumerate}

\newcommand{\defn}[1]{\emph{\textbf{#1}}}

\newcommand{\myparagraph}[1]{\smallskip\noindent {\bf #1.}}

\newcommand{\id}[1]{\ifmmode\mathit{#1}\else\textit{#1}\fi}
\newcommand{\const}[1]{\ifmmode\mbox{\textc{#1}}\else\textsc{#1}\fi}
\usepackage[aboveskip=2pt,belowskip=0pt]{caption}

\usepackage{balance}
\usepackage{microtype}

\usepackage[most]{tcolorbox}
\tcbuselibrary{breakable}

\newtcolorbox{OuterBox}[1][]{%
    breakable,
    enhanced,
    frame hidden,
    interior hidden,
    left=-5pt,
    right=-5pt,
    top=-5pt,
    float=p,
    boxsep=0pt,
    arc=0pt
#1}%

\newtcolorbox{InnerBox}[1][]{%
    enforce breakable,
    enhanced,
    colback=gray,
    colframe=white,
#1}%

\newenvironment{tbox}{
\vspace{0.2cm}
\begin{tcolorbox}[width=\textwidth,
                  enhanced,
                  boxsep=1pt,
                  left=1pt,
                  right=1pt,
                  top=0.5pt,
                  boxrule=1pt,
                  arc=0pt,
                  colback=white,
                  colframe=black,
                  unbreakable
                  ]
}{
\end{tcolorbox}
}

\newenvironment{mytbox}{
\vspace{0.2cm}
\begin{tcolorbox}[width=\textwidth,
                  enhanced,
                  boxsep=1pt,
                  left=1pt,
                  right=1pt,
                  top=4pt,
                  boxrule=1pt,
                  arc=0pt,
                  colback=white,
                  colframe=black,
                  unbreakable
                  ]
}{
\end{tcolorbox}
}

\newcommand{\tboxhrule}[0]{\vspace{0.1cm} {\color{black} \hrule} \vspace{0.2cm}}

\newenvironment{mytitledtbox}[1]{\begin{mytbox}#1 \tboxhrule}{\end{mytbox}}
\newenvironment{benchmark}[1]{\begin{mytitledtbox}{{#1}}}{\end{mytitledtbox}}

\newenvironment{tboxalg}[1]{\begin{algorithm}\caption{#1}}{\end{algorithm}}

\begin{document}

\title[]{Theoretically Efficient Parallel Graph Algorithms\\ Can Be Fast and Scalable}

  \author{Laxman Dhulipala}
  \affiliation{\institution{Carnegie Mellon University}}
  \email{ldhulipa@cs.cmu.edu}
  \author{Guy E. Blelloch}
  \affiliation{\institution{Carnegie Mellon University}}
  \email{guyb@cs.cmu.edu}
  \author{Julian Shun}
  \affiliation{\institution{MIT CSAIL}}
  \email{jshun@mit.edu}




\begin{abstract}

There has been significant recent interest in parallel graph processing
due to the need to quickly analyze the large graphs available
today. Many graph codes have been designed for distributed memory or
external memory. However, today even the largest publicly-available
real-world graph (the Hyperlink Web graph with over 3.5 billion
vertices and 128 billion edges) can fit in the memory of a single
commodity multicore server.  Nevertheless, most experimental work in
the literature report results on much smaller graphs, and the ones
for the Hyperlink graph use distributed or external memory.
Therefore, it is natural to ask whether we can efficiently solve a
broad class of graph problems on this graph in memory.

This paper shows that
theoretically-efficient parallel graph algorithms can scale to the largest
publicly-available graphs using a single machine with a terabyte of RAM,
processing them in minutes.  We give implementations of theoretically-efficient
parallel algorithms for 20 important graph problems.
We also present the optimizations and techniques that we used in our
implementations, which were crucial in enabling us to process these large
graphs quickly.
We show that the running times of our implementations outperform existing
state-of-the-art implementations on the largest real-world graphs.
For many of the problems that we consider, this is the first time they have been
solved on graphs at this scale.
We have made the implementations developed in this work
publicly-available as the Graph-Based Benchmark Suite (GBBS).
\end{abstract}

\maketitle

\section{Introduction}\label{sec:intro}

\begin{table*}[htbp]
  \footnotesize
  \centering

\scalebox{0.92}{

\hspace*{-4.5em}
\begin{tabular}[!t]{|l|c|c|c|c|c|l|l|}
    \toprule
    \textbf{Problem} & (1) & (72h) & (SU) & \textbf{Alg.}& \textbf{Model} & \textbf{Work} & \textbf{Depth}\\
    \midrule
    Breadth-First Search (BFS)                 & 576    & 8.44  & 68  & --                             & \tsmod{}    & $O(m)$                 & $O(\mathsf{diam}(G) \log n)$ \\
    Integral-Weight SSSP (weighted BFS)        & 3770   & 58.1  & 64  & \cite{dhulipala2017julienne}   & \pwmod{}    & $O(m)$ expected        & $O(\mathsf{diam}(G) \log n)$ w.h.p.$^\dag$ \\
    General-Weight SSSP (Bellman-Ford)         & 4010   & 59.4  & 67  & \cite{CLRS}                    & \pwmod{}    & $O(\mathsf{diam}(G)m)$ & $O(\mathsf{diam}(G) \log n)$ \\
    Single-Source Widest Path (Bellman-Ford)                 & 3210   & 48.4  & 66  & \cite{CLRS}                    & \pwmod{}    & $O(\mathsf{diam}(G)m)$ & $O(\mathsf{diam}(G) \log n)$ \\
    Single-Source Betweenness Centrality (BC)  & 2260   & 37.1  & 60  & \cite{brandes2001faster}       & \famod{}    & $O(m)$                 & $O(\mathsf{diam}(G) \log n)$ \\
    $O(k)$-Spanner                             & 2390   & 36.5  & 65  & \cite{miller2015spanners}      & \tsmod{}    & $O(m)$                 & $O(k\log n)$ w.h.p. \\
    Low-Diameter Decomposition  (LDD)          & 980    & 16.6  & 59  & \cite{miller2013parallel}      & \tsmod{}    & $O(m)$                 & $O(\log^2 n)$ w.h.p. \\
    Connectivity                               & 1640   & 25.0  & 65  & \cite{shun2014practical}       & \tsmod{}    & $O(m)$ expected        & $O(\log^3 n)$ w.h.p.  \\
    Spanning Forest                            & 2420   & 35.8  & 67  & \cite{shun2014practical}       & \tsmod{}    & $O(m)$ expected        & $O(\log^3 n)$ w.h.p.  \\
    Biconnectivity                             & 9860   & 165   & 59  &\cite{tarjan1985efficient}      & \famod{}    & $O(m)$ expected        & $O(\max(\mathsf{diam}(G)\log n, \log^{3} n))$ w.h.p. \\
    Strongly Connected Components (SCC)*       & 8130   & 185   & 43  & \cite{blelloch2016parallelism} & \pwmod{}    & $O(m\log n)$ expected  & $O(\mathsf{diam}(G)\log n)$ w.h.p. \\
    Minimum Spanning Forest (MSF)              & 9520   & 187   & 50  & \cite{zhou2017mst}             & \pwmod{}    & $O(m\log n)$           & $O(\log^2 n)$ \\
    Maximal Independent Set (MIS)              & 2190   & 32.2  & 68  & \cite{blelloch2012greedy}      & \famod{}    & $O(m)$ expected        & $O(\log^2 n)$ w.h.p. \\
    Maximal Matching  (MM)                     & 7150   & 108   & 66  & \cite{blelloch2012greedy}      & \pwmod{}    & $O(m)$ expected        & $O(\log^3 m/\log\log m)$ w.h.p. \\
    Graph Coloring                             & 8920   & 158   & 56  & \cite{hasenplaugh2014ordering} & \famod{}    & $O(m)$                 & $O(\log n + L \log \Delta)$ \\
    Approximate Set Cover                      & 5320   & 90.4  & 58  & \cite{blelloch11manis}         & \pwmod{}    & $O(m)$ expected        & $O(\log^{3} n)$ w.h.p. \\
    $k$-core                                   & 8515   & 184   & 46  & \cite{dhulipala2017julienne}   & \famod{}    & $O(m)$ expected        & $O(\rho \log n)$ w.h.p. \\
    Approximate Densest Subgraph               & 3780   & 51.4  & 73  & \cite{bahmani2012densest}      & \famod{}    & $O(m)$                 & $O(\log^2 n)$ \\
    Triangle Counting (TC)                     & ---    & 1168  & --- & \cite{shun2015multicore}       & --     & $O(m^{3/2})$           & $O(\log n)$ \\
    PageRank Iteration                         & 973    & 13.1  & 74 & \cite{brin1998pagerank}        & \famod{}    & $O(n+m)$              & $O(\log n)$ \\
    \bottomrule
  \end{tabular}
}
  \par
    \caption{\footnotesize Running times (in seconds) of our algorithms
      on the symmetrized Hyperlink2012 graph where (1) is the single-thread time,
      (72h) is the 72-core time using hyper-threading, and (SU) is the
      parallel speedup. Theoretical bounds for the algorithms and the
      variant of the \mpram{} used ({\bf MM}) are shown in the last three
      columns. We mark times that did not finish in 5 hours with ---.
      *SCC was run on the directed version of the graph.
      $^\dag$We say that an algorithm has $O(f(n))$ cost \defn{with high probability (w.h.p.)} if it
       has $O(k\cdot f(n))$ cost with probability at least $1 -
       1/n^{k}$.
    }
    \label{table:algorithmcosts}

\end{table*}

Today, the largest publicly-available graph, the Hyperlink Web graph,
consists of over 3.5 billion vertices and 128 billion
edges~\cite{meusel15hyperlink}. This graph presents a significant
computational challenge for both distributed and shared memory
systems.  Indeed, very few algorithms have been applied to this graph,
and those that have often take hours to run~\cite{da2015flashgraph,
  maass2017mosaic, jun2017bigsparse}, with the fastest times
requiring between 1--6 minutes using a
supercomputer~\cite{Slota2015,Slota2016}. In this paper, we
show that a wide range of fundamental graph problems can be solved
quickly on this graph, often in minutes, on a single commodity
shared-memory machine with a terabyte of RAM.\footnote{These machines
  are roughly the size of a workstation and can be easily rented
  in the cloud (e.g., on Amazon EC2).}
For example, our $k$-core implementation takes under 3.5 minutes on 72 cores,
whereas Slota et al.~\cite{Slota2016} report a running time of about 6 minutes for
\emph{approximate} $k$-core on a supercomputer with over 8000 cores.
They also report that they can identify the largest connected
component on this graph in 63 seconds, whereas we can identify
\emph{all} connected components in 25 seconds. Another recent result
by Stergiou et al.~\cite{Stergiou2018} solves connectivity on the
Hyperlink 2012 graph in 341 seconds on a 1000 node cluster with 12000
cores and 128TB of RAM. Compared to this result, our implementation is
13.6x faster on a system with 128x less memory and 166x fewer cores.
However, we note that they are able to process a significantly larger
private graph that we would not be able to fit into our memory
footprint.
A more complete comparison between our work and existing work,
including both distributed and disk-based
systems~\cite{da2015flashgraph, maass2017mosaic, jun2017bigsparse,
  dathathri2018gluon, hoang2019round}, is given in
Section~\ref{sec:exps}.

Importantly, all of our implementations have strong theoretical bounds
on their work and depth.  There are several reasons that algorithms
with good theoretical guarantees are desirable.  For one, they are
robust as even adversarially-chosen inputs will not cause them to
perform extremely poorly.  Furthermore, they can be designed on
pen-and-paper by exploiting properties of the problem instead of
tailoring solutions to the particular dataset at hand.  Theoretical
guarantees also make it likely that the algorithm will continue to
perform well even if the underlying data changes.  Finally,
careful implementations of algorithms that are nearly work-efficient
can perform much less work in practice than work-inefficient
algorithms.  This reduction in work often translates to faster running
times on the same number of cores~\cite{dhulipala2017julienne}.  We
note that most running times that have been reported in the literature on
the Hyperlink Web graph use parallel algorithms that are not
theoretically-efficient.

In this paper, we present implementations of parallel algorithms with
strong theoretical bounds on their work and depth for connectivity, biconnectivity, strongly connected components,
low-diameter decomposition, graph spanners, maximal independent set, maximal matching,
graph coloring, breadth-first search, single-source shortest paths, widest (bottleneck) path,
betweenness centrality, PageRank, spanning forest, minimum spanning forest, $k$-core
decomposition, approximate set cover, approximate densest subgraph, and triangle counting.
We describe the techniques used to achieve good performance on graphs with
billions of vertices and hundreds of billions of edges and share experimental
results for the Hyperlink 2012 and Hyperlink 2014 Web crawls, the largest and
second largest publicly-available graphs, as well as several smaller real-world
graphs at various scales. Some of the algorithms we describe are based on
previous results from Ligra, Ligra+, and Julienne~\cite{shun2012ligra,
  shun2015ligraplus, dhulipala2017julienne}, and other papers on efficient
parallel graph algorithms~\cite{blelloch2012greedy, hasenplaugh2014ordering,
  shun2015multicore}. However, most existing implementations were changed
significantly in order to be more memory efficient. Several algorithm
implementations for problems like strongly connected components, minimum
spanning forest, and biconnectivity are new, and required implementation
techniques to scale that we believe are of independent interest.
We also had to extend the compressed representation from
Ligra+~\cite{shun2015ligraplus} to ensure that our graph
primitives for mapping, filtering, reducing and packing the neighbors of a
vertex were theoretically-efficient. We note that using compression
techniques is crucial for representing the symmetrized Hyperlink 2012
graph in 1TB of RAM, as storing this graph in an
uncompressed format would require over 900GB to store the edges alone,
whereas the graph requires 330GB in our compressed format (less than
1.5 bytes per edge). We show the running times of our algorithms on
the Hyperlink 2012 graph as well as their work and depth bounds in
Table~\ref{table:algorithmcosts}. To make it easy to build upon or
compare to our work in the future, we describe a benchmark suite
containing our problems with clear I/O specifications, which we have
made publicly-available.\footnote{\url{https://github.com/ldhulipala/gbbs}}

We present an experimental evaluation of all of our implementations, and in
almost all cases, the numbers we report are faster than any previous performance
numbers for any machines, even much larger supercomputers. We are also able
to apply our algorithms to the largest publicly-available graph, in many cases
for the first time in the literature, using a reasonably modest machine. Most
importantly, our implementations are based on reasonably simple
algorithms with strong bounds on their work and depth. We believe that
our implementations are likely to scale to larger graphs and lead to
efficient algorithms for related problems.

\section{Related Work}\label{sec:relwork}
\myparagraph{Parallel Graph Algorithms}
Parallel graph algorithms have received significant attention since the start
of parallel computing, and many elegant algorithms with good theoretical bounds have been developed
over the decades (e.g., \cite{shiloach1982connectivity, karp1984mis, luby1986mis, alon1986mis, tarjan1985efficient, Miller1992, Ramachandran89, JaJa92, cole96minimum, pettie2000mst, miller2013parallel, fineman2017reachability,Birn13,meyer2003delta}).
A major goal in parallel graph algorithm design is to find \emph{work-efficient}
algorithms with polylogarithmic depth.
While many suspect
that work-efficient algorithms may not exist for all parallelizable graph
problems, as inefficiency may be inevitable for problems that depend on
transitive closure, many problems that are of practical importance do
admit work-efficient algorithms~\cite{Karp1991}. For these problems,
which include connectivity, biconnectivity, minimum spanning forest, maximal independent set, maximal matching, and
triangle counting, giving theoretically-efficient implementations that are
simple and practical is important, as the amount of parallelism available on
modern systems is still modest enough that reducing the amount of work done is
critical for achieving good performance.
Aside from intellectual curiosity, investigating whether theoretically-efficient graph algorithms also perform well in practice is important, as
theoretically-efficient algorithms are less vulnerable to adversarial inputs
than ad-hoc algorithms that happen to work well in practice.

Unfortunately, some problems that are not known to admit
work-efficient parallel algorithms due to the transitive-closure
bottleneck~\cite{Karp1991}, such as strongly connected components (SCC) and single-source
shortest paths (SSSP) are still important in practice. One method for
circumventing the bottleneck is to give work-efficient algorithms for
these problems that run in depth proportional to the diameter of the
graph---as real-world graphs have low diameter, and theoretical
models of real-world graphs predict a logarithmic diameter, these
algorithms offer theoretical guarantees in
practice~\cite{schudy2008finding, blelloch2016parallelism}. Other
problems, like $k$-core are
$\mathsf{P}$-complete~\cite{anderson84pcomplete}, which rules out
polylogarithmic-depth algorithms for them unless
$\mathsf{P}=\NC$~\cite{Greenlaw1995}. However, even $k$-core admits an
algorithm with strong theoretical guarantees that is efficient in
practice~\cite{dhulipala2017julienne}.

\myparagraph{Parallel Graph Processing Frameworks}
Motivated by the need to process very large graphs, there have been many graph processing frameworks developed in the literature (e.g., \cite{malewicz10pregel,gonzalez2012powergraph,low2010graphlab,Nguyen2013,shun2012ligra} among many others). We refer the reader to~\cite{McCune2015,Yan2017} for surveys of existing frameworks.
Several recent graph processing systems evaluate the scalability of their
implementations by solving problems on massive graphs~\cite{Slota2015,
da2015flashgraph, maass2017mosaic, dhulipala2017julienne, jun2017bigsparse, Stergiou2018}.
All of these systems report running times either on the
Hyperlink 2012 graph or Hyperlink 2014 graphs, two web crawls released by the
WebDataCommons that are the largest and second largest publicly-available
graphs respectively. We describe these recent systems and give a detailed
comparison of how our implementations perform compare to their codes in
Section~\ref{sec:exps}.

\myparagraph{Benchmarking Parallel Graph Algorithms}
There are a surprising number of existing benchmarks of parallel graph
algorithms.  SSCA~\cite{bader2005design}, an early graph processing
benchmark, specifies four graph kernels, which include generating
graphs in adjacency list format, subgraph extraction, and graph
clustering. The Problem Based Benchmark Suite
(PBBS)~\cite{shun2012brief} is a general benchmark of parallel
algorithms that includes 6 problems on graphs (BFS, spanning forest,
minimum spanning forest, maximal independent set, maximal matching,
and graph separators). The PBBS benchmarks are problem-based in that
they are defined only in terms of the input and output without any
specification of the algorithm used to solve the problem.  We follow
the style of PBBS in this paper of defining the input and output
requirements for each problem. The Graph Algorithm Platform (GAP)
Benchmark Suite~\cite{BeamerAP15} specifies 6 kernels for BFS, SSSP,
PageRank, connectivity, betweenness centrality, and triangle counting.

Several recent benchmarks characterize the architectural properties of
parallel graph algorithms. GraphBIG~\cite{nai2015graphbig} describes
12 applications, including several problems that we consider, like
$k$-core and graph coloring (using the Jones-Plassmann algorithm), but
also problems like depth-first search, which are difficult to
parallelize, as well as dynamic graph operations.
CRONO~\cite{ahmad2015crono} implements 10 graph algorithms, including
all-pairs shortest paths, exact betweenness centrality, traveling
salesman, and depth-first search. LDBC~\cite{iosup16ldbc} is an
industry-driven benchmark that selects 6 algorithms that are
considered representative of graph processing including BFS, and
several algorithms based on label propagation.

Unfortunately, all of the existing graph algorithm benchmarks we are
aware of restrict their evaluation to small graphs, often on the order
of tens or hundreds of millions of edges, with the largest graphs in
the benchmarks having about two billion edges. As real-world graphs
are frequently several orders of magnitude larger than this,
evaluation on such small graphs makes it hard to judge whether the
algorithms or results from a benchmark scale to terabyte-scale graphs.

\section{Preliminaries}\label{sec:prelims}
\myparagraph{Graph Notation} We denote an unweighted graph by
$G(V, E)$, where $V$ is the set of vertices and $E$ is the set of
edges in the graph. A weighted graph is denoted by $G = (V,
E, w)$, where $w$ is a function which maps an edge to a real value
(its weight). The number of vertices in a graph is $n = |V|$, and the
number of edges is $m = |E|$. Vertices are assumed to be indexed from
$0$ to $n-1$.  For undirected graphs we use $N(v)$ to denote the
neighbors of vertex $v$ and $\emph{deg}(v)$ to denote its degree. For
directed graphs, we use $\emph{in-deg}(v)$ and $\emph{out-deg}(v)$ to
denote the in and out-neighbors of a vertex $v$. We use
$\mathsf{diam}(G)$ to refer to the diameter of the graph, or the
longest shortest path distance between any vertex $s$ and any vertex
$v$ reachable from $s$.
Given an undirected graph $G=(V, E)$ the density of a set $S \subseteq
V$, or $\rho(S)$, is equal to $\frac{|E(S)|}{|S|}$ where $E(S)$ are
the edges in the induced subgraph on $S$.
$\Delta$ is used to denote the maximum degree
of the graph. We assume that there are no self-edges or duplicate
edges in the graph.  We refer to graphs stored as a list of edges as
being stored in the \defn{edgelist} format and the compressed-sparse
column and compressed-sparse row formats as \defn{CSC} and \defn{CSR}
respectively.

\myparagraph{Atomic Primitives}
We use three common atomic
primitives in our algorithms: test-and-set (TS), fetch-and-add (FA),
and priority-write (PW). A $\textproc{test-and-set}(\&x)$
checks if $x$ is $0$, and if so atomically sets it to $1$ and returns
$\codevar{true}$; otherwise it returns $\codevar{false}$.  A
$\textproc{fetch-and-add}(\&x)$ atomically returns the
current value of $x$ and then increments $x$.  A
$\textproc{priority-write}(\&x, v, p)$ atomically compares
$v$ with the current value of $x$ using the priority function $p$, and
if $v$ has higher priority than the value of $x$ according to $p$ it
sets $x$ to $v$ and returns $\codevar{true}$; otherwise it returns
$\codevar{false}$.

\myparagraph{Model}
In the analysis of algorithms we use the following work-depth model,
which is closely related to the \pram{} but better models current
machines and programming paradigms that are asynchronous and allow
dynamic forking.
We can simulate the model on the \crcwpram{} equipped with the same
operations with an additional $O(\log^{*} n)$ factor in the depth due to
load-balancing. Furthermore, a \pram{} algorithm using $P$ processors and
$T$ time can be simulated in our model with $PT$ work and $T$ depth.

\newcommand{\forkins}{\texttt{fork}}
\newcommand{\insend}{\texttt{end}}
\newcommand{\allocateins}{\texttt{allocate}}
\newcommand{\freeins}{\texttt{free}}

The \emph{Threaded Random-Access Machine} (\mpram{})~\cite{blelloch18notes} consists of a set of
\processes{} that share an unbounded memory.  Each \process{} is basically
equivalent to a Random Access Machine---it works on a program stored
in memory, has a constant number of registers, and has standard \ram{}
instructions (including an \insend{} to finish the computation).  The
\mpram{} extends the \ram{} with a \forkins{} instruction that takes a positive
integer $k$ and forks $k$ new child \processes{}.  Each child \process{}
receives a unique integer in the range $[1,\ldots,k]$ in its first
register and otherwise has the identical state as the parent, which
has a $0$ in that register.  They all start by running the next
instruction.  When a \process{} performs a \forkins{}, it is suspended until
all the children terminate (execute an \insend{} instruction).  A
computation starts with a single root \process{} and finishes when that
root \process{} ends.  This model supports what is often referred to as
nested parallelism.  If the root \process{} never does a fork, it is a
standard sequential program.

A computation can be viewed as a series-parallel DAG in which each
instruction is a vertex, sequential instructions are composed in
series, and the forked sub\processes{} are composed in parallel. The \defn{work}
of a computation is the number of vertices and the \defn{depth} is the length
of the longest path in the DAG.
We augment the model with three atomic instructions that are used
by our algorithms: test-and-set (\tsmod{}), fetch-and-add (\famod{}), and priority-write
(\pwmod{}) and discuss our model with these operations as the \tsmod{},
\famod{}, and \pwmod{} variants of the \mpram{}. As is standard with the
\ram{} model, we assume that the memory locations and registers have
at most $O(\log M)$ bits, where $M$ is the total size of the memory
used. More details about the model can be found
in~\cite{blelloch18notes}.

\myparagraph{Parallel Primitives}
The following parallel procedures are used throughout the paper.
\defn{Scan} takes as input an array $A$ of length $n$, an associative
binary operator $\oplus$, and an identity element $\bot$ such that
$\bot \oplus x = x$ for any $x$, and returns the array
$(\bot, \bot \oplus A[0], \bot \oplus A[0] \oplus A[1], \ldots, \bot \oplus_{i=0}^{n-2} A[i])$
as well as the overall sum, $\bot \oplus_{i=0}^{n-1} A[i]$.
Scan can be done in $O(n)$ work and $O(\log n)$ depth (assuming $\oplus$
takes $O(1)$ work)~\cite{JaJa92}.
\defn{Reduce} takes an array $A$ and a binary associative function
$f$ and returns the sum of the elements in $A$ with respect to $f$.
\defn{Filter} takes an array $A$ and a predicate $f$ and returns a new array
containing $a \in A$ for which $f(a)$ is true, in the same order as in $A$.
Reduce and filter can both be done in $O(n)$ work and $O(\log n)$ depth
(assuming $f$ takes $O(1)$ work).

\myparagraph{Ligra, Ligra+, and Julienne}
We make use of the Ligra, Ligra+, and Julienne frameworks for shared-memory
graph processing in this paper and review components from these frameworks
here~\cite{shun2012ligra, shun2015ligraplus, dhulipala2017julienne}. Ligra
provides data structures for representing a graph $G=(V,E)$,
\defn{\vset{}}s (subsets of the vertices).
We make use of the \emap{} function provided by Ligra, which we use
for mapping over edges.

\emap{} takes as input a graph $G(V,E)$, a \vset{} $U$, and two
boolean functions $F$ and $C$. \emap{} applies $F$ to $(u,v) \in E$
such that $u \in U$ and $C(v) = \codevar{true}$ (call this subset of
edges $E_{a}$), and returns a \vset{} $U'$ where $u \in U'$ if and only
if $(u,v) \in E_{a}$ and $F(u,v) = \codevar{true}$.
$F$ can side-effect data structures associated with the vertices. \emap{}
runs in $O(\sum_{u\in U}\emph{deg}(u))$ work and $O(\log n)$ depth assuming $F$ and $C$
take $O(1)$ work. \emap{} either applies a \defn{sparse} or
\defn{dense} method based on the number of edges incident to the
current frontier. Both methods run in
$O(\sum_{u\in U}\emph{deg}(u))$ work and $O(\log n)$ depth. We note
that in our experiments we use an optimized version of the dense method
which examines in-edges sequentially and stops once $C$ returns $\codevar{false}$. This optimization lets us
potentially examine significantly fewer edges than the $O(\log n)$
depth version, but at the cost of $O(\emph{in-deg}(v))$ depth.

\section{Benchmark}\label{sec:benchmark}

In this section we describe I/O specifications of our benchmark. We
discuss related work and present the theoretically-efficient algorithm
implemented for each problem in Section~\ref{sec:algs}. We mark
implementations based on prior work with a $\dagger$.

\subsection{Shortest Path Problems}\label{sec:sssp}
\begin{benchmark}{Breadth-First Search (BFS)$^{\dagger}$}
Input: $G=(V, E)$, an unweighted graph, $\src \in V$.\\
Output: $D$, a mapping where $D[v]$ is the shortest path distance from $\src$ to $v$ in $G$ and $\infty$ if $v$ is unreachable.
\end{benchmark}

\vspace{-1em}
\begin{benchmark}{Integral-Weight SSSP (weighted BFS)$^{\dagger}$}
Input: $G=(V, E, w)$, a weighted graph with integral edge weights, $\src \in V$.\\
Output: $D$, a mapping where $D[v]$ is the shortest path distance from $\src$ to $v$ in $G$ and $\infty$ if $v$ is unreachable.
\end{benchmark}

\vspace{-1em}
\begin{benchmark}{General-Weight SSSP (Bellman-Ford)$^{\dagger}$}
Input: $G=(V, E, w)$, a weighted graph, $\src \in V$.\\
Output: $D$, a mapping where $D[v]$ is the shortest path distance from $\src$ to $v$ in $G$ and $\infty$ if $v$ is unreachable. All distances must be $-\infty$ if $G$ contains a negative-weight cycle reachable from $\src$.
\end{benchmark}

\vspace{-1em}
\begin{benchmark}{Single-Source Betweenness Centrality (BC)$^{\dagger}$}
Input: $G=(V, E)$, an undirected graph, $\src \in V$.\\
Output: $S$, a mapping from each vertex $v$ to the centrality contribution from all $(\src, t)$ shortest paths that pass through $v$.
\end{benchmark}

\vspace{-1em}
\begin{benchmark}{Widest Path}
Input: $G=(V, E, w)$, a weighted graph with integral edge weights, $\src \in V$.\\
Output: $D$, a mapping where $D[v]$ is the maximum over all paths
between $\src$ and $v$ in $G$ of the minimum weight on the path and
$\infty$ if $v$ is unreachable.
\end{benchmark}

\vspace{-1em}
\begin{benchmark}{$O(k)$-Spanner}
Input: $G=(V, E)$, an undirected, unweighted graph, and an integer
stretch factor, $k$.\\
Output: $H \subseteq E$, a set of edges such that for every $u,v \in V$
connected in $G$, $\dist_{H}(u, v) \leq O(k)$.
\end{benchmark}

\subsection{Connectivity Problems}\label{sec:connectivityproblems}
\begin{benchmark}{Low-Diameter Decomposition$^{\dagger}$}
{Input:} $G=(V, E)$, a directed graph, $0 < \beta < 1$.\\
{Output:} $\mathcal{L}$, a mapping from each vertex to a cluster ID representing a $(O(\beta), O((\log n)/\beta))$ decomposition.
A $(\beta, d)$-decomposition partitions $V$ into $V_{1}, \ldots, V_{k}$ such that
the shortest path between two vertices in $V_{i}$ using only vertices in
$V_{i}$ is at most $d$, and the number of edges $(u,v)$ where
$u \in V_{i}, v \in V_{j}, j \neq i$ is at most $\beta m$.
\end{benchmark}

\vspace{-1em}
\begin{benchmark}{Connectivity$^{\dagger}$}
Input: $G=(V,E)$, an undirected graph.\\
Output: $\mathcal{L}$, a mapping from each vertex to a unique label for its connected component.
\end{benchmark}

\vspace{-1em}
\begin{benchmark}{Spanning Forest$^{\dagger}$}
Input: $G=(V,E)$, an undirected graph.\\
Output: $T$, a set of edges representing a spanning forest of $G$.
\end{benchmark}

\vspace{-1em}
\begin{benchmark}{Biconnectivity}
Input: $G=(V,E)$, an undirected graph.\\
Output: $\mathcal{L}$, a mapping from each edge to the label of its biconnected component.
\end{benchmark}

\vspace{-1em}
\begin{benchmark}{Minimum Spanning Forest}
Input: $G=(V, E, w)$, a weighted graph.\\
Output: $T$, a set of edges representing a minimum spanning forest of $G$.
\end{benchmark}

\vspace{-1em}
\begin{benchmark}{Strongly Connected Components}
Input: $G(V, E)$, a directed graph.\\
Output: $\mathcal{L}$, a mapping from each vertex to the label of its strongly connected component.
\end{benchmark}

\subsection{Covering Problems}\label{sec:coveringproblems}
\begin{benchmark}{Maximal Independent Set$^{\dagger}$}
Input: $G=(V, E)$, an undirected graph.\\
Output: $U \subseteq V$, a set of vertices such that no two vertices in $U$ are neighbors and all vertices in $V \setminus U$ have a neighbor in $U$.
\end{benchmark}

\vspace{-1em}
\begin{benchmark}{Maximal Matching$^{\dagger}$}
Input: $G=(V, E)$, an undirected graph.\\
Output: $E' \subseteq E$, a set of edges such that no two edges in $E'$ share an endpoint and all edges in $E \setminus E'$ share an endpoint with some edge in $E'$.
\end{benchmark}

\vspace{-1em}
\begin{benchmark}{Graph Coloring$^{\dagger}$}
Input: $G=(V, E)$, an undirected graph.\\
Output: $C$, a mapping from each vertex to a color such that for each
edge $(u, v) \in E$, $C(u) \neq C(v)$, using at most $\Delta+1$ colors.
\end{benchmark}

\vspace{-1em}
\begin{benchmark}{Approximate Set Cover$^{\dagger}$}
Input: $G=(V, E)$, an undirected graph representing a set cover instance.\\
Output: $S \subseteq V$, a set of sets such that $\cup_{s \in s} N(s) = V$ with
$|S|$ being an $O(\log n)$-approximation to the optimal cover.
\end{benchmark}

\subsection{Substructure Problems}\label{sec:substructureproblems}
\begin{benchmark}{$k$-core$^{\dagger}$}
Input: $G=(V, E)$, an undirected graph.\\
Output: $D$, a mapping from each vertex to its coreness value.
\end{benchmark}

\vspace{-1em}
\begin{benchmark}{Approximate Densest Subgraph}
Input: $G=(V, E)$, an undirected graph, and a parameter $\epsilon$.\\
Output: $U \subseteq V$, a set of vertices such that the density of $G_{U}$
is a $2(1+\epsilon)$ approximation of density of the densest subgraph of $G$.
\end{benchmark}

\vspace{-1em}
\begin{benchmark}{Triangle Counting$^{\dagger}$}
Input: $G=(V, E)$, an undirected graph.\\
Output: $T_{G}$, the total number of triangles in $G$.
\end{benchmark}

\subsection{Eigenvector Problems}\label{sec:eigenvectorproblems}
\begin{benchmark}{PageRank$^{\dagger}$}
Input: $G=(V, E)$, an undirected graph.\\
Output: $\mathcal{P}$, a mapping from each vertex to its PageRank
value after a single iteration of PageRank.
\end{benchmark}

\section{Algorithms}\label{sec:algs}

In this section we discuss related work and give self-contained
descriptions of all of the theoretically efficient algorithms
implemented in our benchmark.  We also include implementation details
for algorithms implemented in prior papers that we did not
significantly change. The pseudocode for many of the algorithms make
use of the \emap{} and \vmap{} primitives, as well as test-and-set,
fetch-and-add and priority-write. All of these primitives are defined
in Section~\ref{sec:prelims}.  We cite the original papers that our
algorithms are based on in Table~\ref{table:algorithmcosts}.

\begin{tboxalg}{Breadth-First Search} \label{alg:bfs}
\small
\begin{algorithmic}[1]
\State $\codevar{Fl} = \{0, \ldots, 0\}$,\ $\codevar{D} = \{\infty, \ldots, \infty\}$,\ $\codevar{round} = 0$
\Procedure{Cond}{$v$} \algorithmicreturn{} $\codevar{Fl}[v] == 0$
\EndProcedure
\Procedure{Update}{$s$, $d$}
\If{ ($!\codevar{Fl}[d]$ \&\& $(\textproc{\ts{}}(\&Fl[d]) == 0)$) }
\State $\codevar{D}[d] = \codevar{round}$
\State \algorithmicreturn{} 1
\EndIf
\State \algorithmicreturn{} 0
\EndProcedure
\Procedure{BFS}{$G(V, E), r$}
\State $F = \{ r\}$,\ $D[r] = 0$
\While {$|F| > 0$}
  \State $F = \textproc{edgeMap}(G, F, \textproc{Update}, \textproc{Cond})$
  \State $\codevar{round} = \codevar{round} + 1$
\EndWhile
\State \algorithmicreturn{} $\codevar{D}$
\EndProcedure
\end{algorithmic}
\end{tboxalg}

\begin{tboxalg}{Betweenness Centrality} \label{alg:betweenness}
\small
\begin{algorithmic}[1]
\State $\codevar{Visited} = \{0, \ldots, 0\}$,\ $\codevar{NumPaths} = \{0, \ldots, 0\}$,\ $\codevar{D} = \{0, \ldots, 0\}$
\State $\codevar{round} = 0$,\ $\codevar{Levels} = \mathsf{array}(n, \mathsf{null})$
\Procedure{Cond}{$v$} \algorithmicreturn{} $\codevar{Visited}[v] == 0$
\EndProcedure
\Procedure{SetVisited}{$v$} \algorithmicreturn{} $\codevar{Visited}[v] = 1$
\EndProcedure

\Procedure{PathUpdate}{$s$, $d$}
  \State $\codevar{prev} = \textproc{\fa}(\&\codevar{NumPaths}[d], \codevar{NumPaths}[s])$
  \State \algorithmicreturn{} $\codevar{prev} == 0$
\EndProcedure

\Procedure{DependencyUpdate}{$s$, $d$}
  \State $\codevar{add\_val} = (\codevar{NumPaths}[d] / \codevar{NumPaths}[s]) \cdot (1 + D[s])$
  \State $\textproc{\fa}(\&D[d], \codevar{add\_val})$
\EndProcedure

\Procedure{BC}{$G(V, E), r$}
\State $F = \{ r\}$
\While {$|F| > 0$}
  \State $F = \emap{}(G, F, \textproc{PathUpdate}, \textproc{Cond})$
  \State $\codevar{Levels}[\codevar{round}] = F$
  \State $\vmap{}(F, \textproc{SetVisited})$
  \State $\codevar{round} = \codevar{round} + 1$
\EndWhile
\State $\codevar{Visited} = \{0, \ldots, 0 \}$
\While {$\codevar{round} > 0$}
  \State $F = \codevar{Levels}[\codevar{round} - 1]$
  \State $\vmap{}(F, \textproc{SetVisited})$
  \State $\emap{}(G, F, \textproc{DependencyUpdate}, \textproc{Cond})$
  \State $\codevar{round} = \codevar{round} - 1$
\EndWhile
\State \algorithmicreturn{} $\codevar{D}$
\EndProcedure
\end{algorithmic}
\end{tboxalg}

\begin{tboxalg}{wBFS} \label{alg:wBFS}
\small
\begin{algorithmic}[1]
\State $D = \{\infty,\ldots,\infty\}$,\ $Fl = \{0,\ldots,0\}$
\Procedure{GetBucketNum}{$i$} \Return $D[i]$
\EndProcedure
\Procedure{Cond}{$v$} \algorithmicreturn{} $1$
\EndProcedure
\Procedure{Update}{$s$, $d$, $w$}
\State $\codevar{nDist} = D[s] + w,\ \codevar{oDist} = D[d],\ \codevar{res} = \textproc{None}$
\If{ ($\codevar{nDist} < \codevar{oDist}$) }
  \If{ ($\textproc{\ts{}}(\&Fl[d]) == 0$) } $\codevar{res} = \textproc{Some}(\codevar{oDist})$
  \EndIf
  \State $\textproc{\writemin{}}(\&D[d], \codevar{nDist})$
\EndIf
\State \Return $\codevar{res}$ 
\EndProcedure

\Procedure{Reset}{$v$, $\codevar{oldDist}$}
  \State $Fl[v] = 0,\ \codevar{newDist} = D[d]$
  \State \Return $\codevar{B}.\textproc{get\_bucket}(\codevar{oldDist}, \codevar{newDist})$
\EndProcedure

\Procedure{wBFS}{$G(V, E, W), r$}
\State $D[r] = 0$
\State $\codevar{B} = \makebkt{}(|V|, \textproc{GetBucketNum}, \codevar{\orderinc{}})$
\While { ($(\codevar{id}, \codevar{ids}) = \codevar{B}.\nextbkt{}() \text{ and } \codevar{id} \neq \nullbkt{}$) }
  \State $\codevar{Moved} = \textproc{edgeMapData}(G, \codevar{ids}, \textproc{Update}, \textproc{Cond})$
  \State $\codevar{NewBuckets} = \textproc{vertexMap}(\codevar{Moved}, \textproc{Reset})$
  \State $\codevar{B}$.\updatebkt{}($\codevar{NewBuckets}$, $\vert \codevar{NewBuckets} \vert$)
\EndWhile
\State \algorithmicreturn{} $D$
\EndProcedure
\end{algorithmic}
\end{tboxalg}

\begin{tboxalg}{Bellman-Ford} \label{alg:bellman-ford}
\small
\begin{algorithmic}[1]
\State $\codevar{Fl} = \{0, \ldots, 0\}$,\ $\codevar{D} = \{\infty, \ldots, \infty\}$
\Procedure{Cond}{$v$} \algorithmicreturn{} $1$
\EndProcedure
\Procedure{ResetFlags}{$v$} $\codevar{Fl}[v] = 0$
\EndProcedure
\Procedure{Update}{$s$, $d$, $w$}
\If{ $D[s] + w < D[d]$ }
  \State $\textproc{\writemin}(\&D[d], D[s] + w)$
  \If{ $!\codevar{Fl}[d]$ }
    \algorithmicreturn{} $\textproc{\ts{}}(\&Fl[d]) == 0$
  \EndIf
\EndIf
\State \algorithmicreturn{} 0
\EndProcedure
\Procedure{BellmanFord}{$G(V, E, W), r$}
\State $F = \{ r\}$,\ $D[r] = 0$
\While {$|F| > 0$}
  \State $F = \emap{}(G, F, \textproc{Update}, \textproc{Cond})$
  \State $\vmap{}(F, \textproc{ResetFlags})$
\EndWhile
\State \algorithmicreturn{} $\codevar{D}$
\EndProcedure
\end{algorithmic}
\end{tboxalg}

\subsection{Shortest Path Problems}\label{subsec:sssp}
Although work-efficient polylogarithmic-depth algorithms for single-source
shortest paths (SSSP) type problems are not known due to the transitive-closure
bottleneck~\cite{Karp1991}, work-efficient algorithms that run in depth
proportional to the diameter of the graph are known for the
special cases considered in our benchmark. Several work-efficient parallel
breadth-first search algorithms are
known~\cite{bader2006bfs,leiserson2010bfs,BFGS}.  On weighted graphs with
integral edge weights, SSSP can be solved in $O(m)$ work and
$O(\mathsf{diam}(G)\log n)$ depth~\cite{dhulipala2017julienne}. Parallel
algorithms also exist for weighted graphs with positive edge
weights~\cite{meyer2003delta, meyer2000parallel}. SSSP on graphs with negative
integer edge weights can be solved using Bellman-Ford~\cite{CLRS}, where the
number of iterations depends on the diameter of the graph. Betweenness
centrality from a single source can be computed using two breadth-first
searches~\cite{brandes2001faster, shun2012ligra}.

In this paper, we present implementations of five SSSP problems that
are based on graph search. We also include an algorithm to construct
an $O(k)$-spanner which is based on computing low-diameter
decompositions.

\myparagraph{BFS, wBFS, Bellman-Ford, and Betweenness Centrality}
Our implementations of BFS, Bellman-Ford, and betweenness centrality
are based on the implementations in Ligra~\cite{shun2012ligra}. Our
wBFS implementation is based on our earlier work on
Julienne~\cite{dhulipala2017julienne}.  Algorithm~\ref{alg:bfs} shows
pseudocode for a frontier-based BFS implementation which synchronizes
after each round of the BFS. The algorithm runs in $O(m)$ and
$O(\mathsf{diam}(G)\log n)$ depth on the \tsmod{}-\mpram{}, as
vertices use test-and-set to non-deterministically acquire unvisited
neighbors on the next frontier.  Algorithm~\ref{alg:bellman-ford}
shows pseudocode for a frontier-based version of Bellman-Ford which
uses a priority-write to write the minimum distance to a vertex on
each round and runs in $O(\mathsf{diam}(G)m)$ work and
$O(\mathsf{diam}(G)\log n)$ depth on the \pwmod{}-\mpram{}. Pseudocode
for our betweenness centrality implementation is shown in
Algorithm~\ref{alg:betweenness}. It uses fetch-and-add to compute the
total number of shortest-paths through vertices, and runs in $O(m)$
work and $O(\mathsf{diam}(G)\log n)$ depth on the \famod{}-\mpram{}.
Algorithm~\ref{alg:wBFS} shows pseudocode for our weighted BFS
implementation from Julienne~\cite{dhulipala2017julienne}. The
algorithm runs in $O(m)$ work in expectation and
$O(\mathsf{diam}(G)\log n)$ depth w.h.p on the \pwmod{}-\mpram{}, as
vertices use priority-write to write the minimum distance to a
neighboring vertex on each round. The algorithm uses a bucketing
structure and generalized version of \emap{} called \emapdata{}, which
are discussed in the Julienne paper. The main change we made to these
algorithms was to improve the cache-efficiency of the \emap{}
implementation using the block-based version of \emap{}, described in
Section~\ref{sec:techniques}.

\myparagraph{Widest Path (Bottleneck Path)} The Widest Path, or
Bottleneck Path problem is to compute $\forall v \in V$ the maximum
over all paths of the minimum weight edge on the path between a source
vertex, $u$, and $v$. The algorithm is an important primitive, used
for example in the Ford-Fulkerson flow
algorithm~\cite{ford2009maximal, CLRS}, as well as other flow
algorithms~\cite{baier2005k}. Sequentially, the algorithm can be
solved as quickly as SSSP using a modified version of Dijkstra's
algorithm. We note that faster algorithms are known sequentially for
sparse graphs~\cite{duan2018bottleneck}. For positive integer-weighted
graphs, the problem can also be solved using the the work-efficient
bucketing data structure from Julienne~\cite{dhulipala2017julienne}.
The buckets represent the width classes. The buckets are initialized
with the out-neighbors of the source, $u$, and the buckets are
traversed using the \emph{decreasing} order.
This order specifies that the buckets are traversed from the largest
bucket to the smallest bucket. Unlike the other applications in
Julienne, using widest path is interesting since the bucket containing
a vertex can only increase. The problem can also be solved using the
Bellman-Ford approach, by performing computations over the
$(\max,\min)$ semi-ring instead of the $(\min, +)$ semi-ring. Other
than these changes, the pseudocode for the problem is identical to
that of Algorithms~\ref{alg:wBFS} and~\ref{alg:bellman-ford}.

\begin{tboxalg}{$O(k)$-Spanner}\label{alg:spanner}
\small
\begin{algorithmic}[1]
\Procedure{Spanner}{$G(V, E), k$}
\State $\beta = \frac{\log n}{2k}$
\State $\codevar{L}$ = \textproc{LDD}($G(V, E), \beta$)\label{spanner:ldd}
\State $\codevar{S} = \codevar{E}_{\codevar{L}} =$ tree edges used in the LDD, $L$ \label{spanner:addtree}
\State For each pair of adjacent clusters in $L$, add one inter-cluster edge to $S$. \label{spanner:addedge}
\State \algorithmicreturn{} $\codevar{S}$
\EndProcedure
\end{algorithmic}
\end{tboxalg}

\myparagraph{$O(k)$-Spanner}
Computing
graph spanners is a fundamental problem in combintaorial graph
algorithms and graph theory~\cite{peleg1989graph}. A graph $H$ is a
$k$-spanner of a graph $G$ if $\forall u,v \in V$ connected by a path,
$\dist{}_{G}(u, v) \leq \dist{}_{H}(u, v) \leq k \cdot \dist{}_{G}(u,
v)$. Sequentially, classic results give elegant constructions of
$(2k-1)$-spanners using $O(n^{1+1/k})$ edges, which are essentially
the best possible assuming the girth
conjecture~\cite{thorup2005approximate}. In this paper, we implement
the recent spanner algorithm proposed by Miller, Peng, Xu,
and Vladu (MPXV)~\cite{miller2015spanners}. The construction results
in an $O(k)$-spanner with expected size $O(n^{1+1/k})$, and runs in
$O(m)$ work and $O(k\log n)$ depth on the \tsmod{}-\mpram{}.

The MPXV spanner algorithm (Algorithm~\ref{alg:spanner}) uses the low-diameter decomposition (LDD) algorithm, which will be described in Section~\ref{subsec:connprobs}. It takes as input a parameter $k$ which controls
the multiplicative stretch of the spanner. The idea is to first
compute a LDD with $\beta = \log n/(2k)$
(Line~\ref{spanner:ldd}). The stretch of each ball is $O(k)$
w.h.p., and so the algorithm includes all spanning tree edges
generated by the LDD in the spanner (Line~\ref{spanner:addtree}).
Next, the algorithm handles inter-cluster edges by taking a single
inter-cluster edge between a boundary vertex and its neighbors (Line~\ref{spanner:addedge}). We note
that this procedure is slightly different than the procedure in the
MPXV paper, which adds a single edge between \emph{every} boundary
vertex of a cluster and each adjacent cluster. Our algorithm only adds
a single edge between two clusters, while the MPXV algorithm may add
multiple parallel edges between two clusters. Their argument bounding
the stretch to $O(k)$ for an edge spanning two clusters is still
valid, since the endpoints can be first routed to the cluster centers,
and then to the single edge that was selected between the two
clusters.

\subsection{Connectivity Problems}\label{subsec:connprobs}

\begin{tboxalg}{Low Diameter Decomposition} \label{alg:ldd}
\small
\begin{algorithmic}[1]
\State $\codevar{\delta} = \{\sim Exp(\beta), \ldots, \sim Exp(\beta)\}$
\State $\codevar{Start} = \{\delta_{\codevar{max}} - \delta_{v} \mid v \in V\}$
\State $\codevar{C} = \{\infty, \ldots, \infty\}$

\Procedure{Update}{$s$, $d$}
\If{ ($\codevar{C}[d] == \infty$) }
  \algorithmicreturn{} $\textproc{\cas{}}(\&C[d], \infty, s)$
\EndIf
\State \algorithmicreturn{} 0
\EndProcedure

\Procedure{LDD}{$G(V, E)$}
\State $\codevar{numVisited} = 0, \codevar{round} = 0$
\State $F = \{ \}$
\While {$\codevar{numVisited} < |V|$}
  \State $F = F \cup \{v \in V \mid \codevar{Start}[v] < round + 1\ \text{and}\ \codevar{C}[v] == \infty\}$
  \State $\codevar{numVisited} = \codevar{numVisited} + |F|$
  \State $F' = \textproc{edgeMap}(G, F, \textproc{Update})$
  \State $\codevar{round} = \codevar{round} + 1$
\EndWhile
\State \algorithmicreturn{} $C$
\EndProcedure
\end{algorithmic}
\end{tboxalg}

\myparagraph{Low-Diameter Decomposition}
Low-diameter decompositions (LDD) were first introduced in the context of distributed
computing~\cite{awerbuch1985complexity}, and were later used in metric
embedding, linear-system solvers, and parallel algorithms. Awerbuch presents a
simple sequential algorithm based on ball growing that computes an $(\beta,
O((\log n)/\beta)$ decomposition~\cite{awerbuch1985complexity}.
Miller, Peng, and Xu (MPX) present a work-efficient parallel algorithm that
computes a $(\beta, O((\log n)/\beta)$ decomposition.
For each $v \in V$, the algorithm draws a start time, $\delta_{v}$, from an
exponential distribution with parameter $\beta$. The clustering is done by
assigning each vertex $u$ to the center $v$ which minimizes $d(u, v) -
\delta_{v}$. This algorithm can be implemented by running a set of parallel
breadth-first searches where the initial breadth-first search starts at the
vertex with the largest start time, $\delta_{\max}$, and starting breadth-first
searches from other $v \in V$ once $\delta_{\max} - \delta_{v}$ steps have
elapsed.
In this paper, we present an implementation of the MPX algorithm
which computes an $(2\beta, O(\log n / \beta))$ decomposition in
$O(m)$ expected work and $O(\log^2 n)$ depth w.h.p. on the
\tsmod{}-\mpram{}.  Our implementation is based on the
non-deterministic LDD implementation from Shun et
al.~\cite{shun2014practical}. The main changes in our implementation
are separating the LDD code from the connectivity implementation.

Algorithm~\ref{alg:ldd} shows pseudocode for the modified version of
the Miller-Peng-Xu algorithm from~\cite{shun2014practical}, which
computes a $(2\beta, O(\log n / \beta))$ decomposition in $O(m)$
expected work and $O(\log^2 n)$ depth w.h.p. on the \tsmod{}-\mpram{}.
The algorithm allows ties to be broken arbitrarily when two searches
visit a vertex in the time-step, and one can show that this only
affects the number of cut edges by a constant
factor~\cite{shun2014practical}.  The algorithm first draws
independent samples from $Exp(\beta)$ (Line 1). Next, it computes the
start times which is the difference between the maximum shifted value
$\delta_{\max}$ and $\delta_{v}$ (Line 2). Initially, all vertices are
unvisited (Line 3). The algorithm performs ball-growing while all of
the vertices are not yet visited (Line 10). On Line 11, it updates the
current frontier with any vertices that are ready to start and have
not yet been visited, and update the number of visited vertices with
the size of the current frontier (Line 12).  Finally, on Line 13, it
uses \emap{} to traverse the out edges of the current frontier and
non-deterministically acquire unvisited neighboring vertices.

\begin{tboxalg}{Connectivity} \label{alg:connectivity}
\small
\begin{algorithmic}[1]
\Procedure{Connectivity}{$G(V, E), \beta$}
\State $\codevar{L}$ = \textproc{LDD}($G(V, E), \beta$)
\State $G'(V', E')$ = \textproc{Contract}($G(V, E), L$)
\If{$|E'| = 0$}
  \State \algorithmicreturn{} $\codevar{L}$
\EndIf
\State $\codevar{L}'$ = \textproc{Connectivity}($G'(V', E'), \beta$)
\State $\codevar{L}''$ = $\{ v \rightarrow \codevar{L}'[\codevar{L}[v]]\ |\ v \in V\}$
\State \algorithmicreturn{} $L''$
\EndProcedure
\end{algorithmic}
\end{tboxalg}

\myparagraph{Connectivity}
Connectivity can be solved sequentially in linear work using
breadth-first or depth-first search. Parallel algorithms for connectivity have a long
history; we refer readers to~\cite{shun2014practical} for a review of the
literature. Early work on parallel connectivity discovered many natural algorithms
which perform $O(m\log n)$ work~\cite{shiloach1982connectivity,
  awerbuch1983connectivity, reif1985cc, phillips1989contraction}.
A number of optimal parallel connectivity algorithms were discovered in subsequent
years~\cite{gazit1991connectivity, cole96minimum, halperin1994cc, halperin2000cc,
  poon1997msf, pettie2000mst, shun2014practical}, but to the best of our
knowledge the recent algorithm by Shun et al. is the only linear-work polylogarithmic-depth parallel algorithm that has been studied experimentally~\cite{shun2014practical}.

In this paper we implement the connectivity algorithm from Shun et
al.~\cite{shun2014practical}, which runs in $O(m)$ expected work and
$O(\log^3 n)$ depth w.h.p. on the \tsmod{}-\mpram{}. The
implementation uses the work-efficient algorithm for low-diameter
decomposition (LDD)~\cite{miller2013parallel} described above.  One
change we made to the implementation from ~\cite{shun2014practical}
was to separate the LDD and contraction steps from the connectivity
algorithm.  Refactoring these sub-routines allowed us to express the
main connectivity algorithm in Ligra in about 50 lines of code.

The connectivity algorithm from Shun et al.~\cite{shun2014practical}
(Algorithm~\ref{alg:connectivity}) takes as input an undirected graph
$G$ and a parameter $0 < \beta < 1$. It first runs the LDD algorithm
(Line 2) which decomposes the graph into clusters each with diameter
$(\log n)/\beta$, and $\beta m$ inter-cluster edges in expectation.
Next, it builds $G'$ by contracting each cluster to a single vertex and
adding inter-cluster edges while removing duplicate edges and isolated
vertices (Line 3). It then checks if the contracted graph consists of
isolated vertices (Line 4); if so, the clusters are the components,
and it returns the mapping from vertices to clusters (Line 5).
Otherwise, it recurses on the contracted graph (Line 7) and returns the
connectivity labeling produced by assigning each vertex to the label
assigned to its cluster in the recursive call (Line 9).



\begin{tboxalg}{Spanning Forest} \label{alg:spanning-forest}
\small
\begin{algorithmic}[1]
\Procedure{SpanningForestImpl}{$G(V, E), M, \beta$}
\State $\codevar{L}$ = \textproc{LDD}($G(V, E), \beta$)
\State $\codevar{E}_{\codevar{L}} =$ edges used in the LDD
\State $\codevar{E}_{M} = \{M(e)\ |\ e \in \codevar{E}_{\codevar{L}}\}$\label{sf:lddedges}
\Comment{Original edges corresponding to $\codevar{E}_{\codevar{L}}$}\label{sf:orig}
\State $G'(V', E')$ = \textproc{Contract}($G(V, E), L$)
\If{$|E'| = 0$}
  \State \algorithmicreturn{} $\codevar{L}$
\EndIf
\State $M' = $ mapping from $e' \in E'$ to $M(e)$ where $e \in E$ is
some edge representing $e'$\label{sf:updateM}
\State $\codevar{E}''$ = \textproc{SpanningForest}($G'(V', E'), M', \beta$)
\State \algorithmicreturn{} $\codevar{E}_{M} \cup \codevar{E}''$
\EndProcedure

\Procedure{SpanningForest}{$G(V, E), \beta$}
\State \algorithmicreturn{} \textsc{SpanningForestImpl}$(G, \{e \rightarrow e\ |\ e \in E\}, \beta)$)\label{sf:callimpl}
\EndProcedure
\end{algorithmic}
\end{tboxalg}

\myparagraph{Spanning Forest}
Finding spanning forests in parallel has been studied largely in
conjunction with connectivity algorithms, since most parallel
connectivity algorithms can naturally be modified to output a spanning
forest (see~\cite{shun2014practical} for a review of the literature).
Our spanning forest algorithm (Algorithm~\ref{alg:spanning-forest}) is
based on the connectivity algorithm from Shun et
al.~\cite{shun2014practical} which we described earlier. The main
difference in the spanning forest algorithm is to include all LDD
edges at each level of the recuursion (Line~\ref{sf:lddedges}).
However, observe that the LDD edges after the topmost level of
recursion are taken from a contracted graph, and need to be mapped
back to some edge in the original graph realizing the contracted edge.
We decide which edges in $G$ to add by maintaining a mapping from the
edges in the current graph at some level of recursion to the original
edge set. Initially this mapping, $M$, is an identity map
(Line~\ref{sf:callimpl}). To compute the mapping to pass to the
recursive call, we select any edge $e$ in the input graph $G$
that resulted in $e' \in E'$ and map $e'$ to $M(e)$
(Line~\ref{sf:updateM}). In our implementation, we use a parallel
hash table to select a single original edge per contracted edge.

\begin{tboxalg}{Biconnectivity} \label{alg:biconnectivity}
\small
\begin{algorithmic}[1]
\Procedure{Biconnectivity}{$G(V, E)$}
\State $\codevar{F}$ = \textproc{SpanningForest}($G$) \Comment{trees in $F$ are rooted arbitrarily}
\State $\codevar{PN}$ = \textproc{PreorderNumber}($F$)
\State For each $v \in V$, compute $\codevar{Low}(v)$ and $\codevar{High}(v)$ and $\codevar{Size}(v)$
\State $\codevar{critical}$ = $e = (u, p(u)) \in F$ s.t. $p(u)$ is an articulation point
\State $\codevar{Labels}$ = \textproc{CC}($G(V, E \setminus \codevar{critical})$)
\State \algorithmicreturn{} $(\codevar{Labels}, F)$ \Comment{sufficient to answer biconnectivity queries}
\EndProcedure
\end{algorithmic}
\end{tboxalg}

\myparagraph{Biconnectivity}
Sequentially, biconnectivity can be solved using the Hopcroft-Tarjan
algorithm~\cite{hopcroft1973algorithm}. The algorithm uses depth-first
search (DFS) to identify articulation points and requires $O(m + n)$
work to label all edges with their biconnectivity label. It is
possible to parallelize the sequential algorithm using a parallel DFS,
however, the fastest parallel DFS algorithm is not
work-efficient~\cite{aggarwal1989parallel}. Tarjan and Vishkin present
the first work-efficient algorithm for
biconnectivity~\cite{tarjan1985efficient} (as stated in the paper the
algorithm is not work-efficient, but it can be made so by using a
work-efficient connectivity algorithm). Another approach relies on the
fact that biconnected graphs admit open ear decompositions to solve
biconnectivity efficiently~\cite{maon1986parallel,
ramachandran1992parallel}.

In this paper, we implement the Tarjan-Vishkin algorithm for
biconnectivity in $O(m)$ expected work and
$O(\max(\mathsf{diam}(G)\log n, \log^3 n))$ depth on the
\famod{}-\mpram{}. Our implementation first computes connectivity
labels using our connectivity algorithm, which runs in $O(m)$ expected
work and $O(\log^3 n)$ depth w.h.p. and picks an arbitrary source
vertex from each component. Next, we compute a spanning forest rooted
at these sources using breadth-first search, which runs in $O(m)$ work
and $O(\mathsf{diam}(G)\log n)$ depth. We note that the connectivity
algorithm can be modified to compute a spanning forest in the same
work and depth as connectivity, which would avoid the
breadth-first-search. We compute $\codevar{Low}$, $\codevar{High}$,
and $\codevar{Size}$ for each vertex by running leaffix and rootfix
sums on the spanning forests produced by BFS with fetch-and-add, which
requires $O(n)$ work and $O(\mathsf{diam}(G))$ depth. Finally, we
compute an implicit representation of the biconnectivity labels for
each edge, using an idea from~\cite{bendavid2017implicit}. This step
computes per-vertex labels by removing all critical edges and
computing connectivity on the remaining graph. The resulting vertex
labels can be used to assign biconnectivity labels to edges by giving
tree edges the connectivity label of the vertex further from the root
in the tree, and assigning non-tree edges the label of either
endpoint.  Summing the cost of each step, the total work of this
algorithm is $O(m)$ in expectation and the total depth is
$O(\max(\mathsf{diam}(G)\log n, \log^3 n))$ w.h.p.

Algorithm~\ref{alg:biconnectivity} shows the Tarjan-Vishkin
biconnectivity algorithm.  We first compute a spanning forest of $G$
using the work-efficient connectivity algorithm, where the trees in
the forest can be rooted arbitrarily (Line 2). Next, we compute a
preorder numbering, $\codevar{PN}$, with respect to the roots (Line
3). We then compute for each $v \in V$ $\codevar{Low}(v)$ and
$\codevar{High}(v)$, which are the minimum and maximum preorder
numbers respectively of all non-tree $(u, w)$ edges where $u$ is a
vertex in $v$'s subtree. We also compute $\codevar{Size}(v)$, the size
of each vertex's subtree. Note that we can determine whether the
parent of a vertex $u$ is an articulation point by checking
$\codevar{PN}(p(u)) \leq \codevar{Low}(u)$ and $\codevar{High}(u) <
\codevar{PN}(p(u)) + \codevar{size}(p(u))$.  As
in~\cite{bendavid2017implicit}, we refer to this set of tree edges
$(u,p(u))$, where $p(u)$ is an articulation point, as \emph{critical
edges}. The last step of the algorithm is to solve connectivity on the
graph with all critical edges removed. Now, the biconnectivity label
of an edge $(u,v)$ is the connectivity label of the vertex that is
further from the root of the tree. The query data structure can thus
report biconnectivity labels of edges in $O(1)$ time using $2n$ space,
which is important for our implementations as storing a biconnectivity
label per-edge explicitly would require a prohibitive amount of memory
for large graphs. As the most costly step in this algorithm is to run
connectivity, the algorithm runs in $O(m)$ work in expectation and
$O(\log^{3} n)$ depth w.h.p. Our implementation described of the
Tarjan-Vishkin algorithm runs in the same work but
$O(\max(\mathsf{diam}(G)\log n, \log^{3}n))$ depth w.h.p. as it
computes a spanning tree using BFS and performs leaffix and rootfix
computations on this tree.


\begin{tboxalg}{Minimum Spanning Forest} \label{alg:msf}
\small
\begin{algorithmic}[1]
\Procedure{\Boruvka}{$n, E=\{(v_{i}, v_{j}, w_{v_{i}v_{j}}), \ldots, \}$}
  \State $\codevar{Parents} = \{0, \ldots, n-1\}$, $\codevar{active} = n$, $\codevar{forest} = \{\}$
  \While{$\codevar{active} > 1 \text{ and } |E| > 0$}
    \State $\codevar{P} = \{(\infty, \infty), \ldots, (\infty, \infty)\}$
    \For{($(i, e = (u, v, w)) \in E$} {\bf in parallel}
      \State $\textproc{writeMin}(\&P[u],(w, i))$
      \State $\textproc{writeMin}(\&P[v],(w, i))$
    \EndFor
    \State $\textproc{MarkRoots}(P, \codevar{Parents})$
    \State $\textproc{FilterInactiveVertices}(active, P)$
    \State $\codevar{forest} = \codevar{forest} \cup \{\text{edges that won on either endpoint in $P$}\}$
    \State $\textproc{PointerJump}(\codevar{Parents})$
    \State $\textproc{RelabelEdges}(E, \codevar{Parents})$
    \State $\textproc{FilterSelfEdges}(E)$
  \EndWhile
  \State\algorithmicreturn{} $\codevar{forest}$
\EndProcedure
\end{algorithmic}
\end{tboxalg}

\myparagraph{Minimum Spanning Forest}
\Boruvka{} gave
the first known sequential and parallel algorithm for computing a
minimum spanning forest (MSF)~\cite{boruvka1926a}.  Significant effort
has gone into finding linear-work MSF algorithms both in the
sequential and parallel settings~\cite{Karger1995,cole96minimum,
  pettie2000mst}.  Unfortunately, the linear-work parallel algorithms
are highly involved and do not seem to be practical.  Significant
effort has also gone into designing practical parallel algorithms for
MSF; we discuss relevant experimental work in
Section~\ref{sec:exps}. Due to the simplicity of \Boruvka{}, many
parallel implementations of MSF use variants of it.

In this paper, we present an implementation of \Boruvka's algorithm that runs in
$O(m\log n)$ work and $O(\log^2 n)$ depth on the \pwmod{}-\mpram{}.  Our implementation
is based on a recent implementation of \Boruvka{} by Zhou~\cite{zhou2017mst} that
runs on the edgelist format. We made several changes to the algorithm which
improve performance and allow us to solve MSF on graphs stored in the CSR/CSC
format, as storing an integer-weighted graph in edgelist format would require
well over 1TB of memory to represent the edges in the Hyperlink2012 graph
alone. Our code uses an implementation
of \Boruvka{} that works over an edgelist; to make it efficient we ensure that
the size of the lists passed to it are much smaller than $m$. Our
approach is to perform a constant number of \emph{filtering}
steps. Each filtering step solves an approximate $k$'th smallest
problem in order to extract the
lightest $3n/2$ edges in the graph (or all remaining edges) and runs \Boruvka{}
on this subset of edges. We then filter the remaining graph, packing out any
edges that are now in the same component. This idea is similar to the
theoretically-efficient algorithm of Cole et al.~\cite{cole96minimum}, except that instead
of randomly sampling edges, we select a linear number of the lowest weight edges.
Each filtering step costs $O(m)$ work and $O(\log m)$ depth, but as we only perform a
constant number of steps, they do not affect the work and depth
asymptotically. In practice, most of the edges are removed
after 3--4 filtering steps, and so the remaining edges can be
copied into an edgelist and solved in a single \Boruvka{} step.  We
also note that as the edges are initially represented in both
directions, we can pack out the edges so that each undirected edge is
only inspected once (we noticed that earlier edgelist-based
implementations stored undirected edges in both directions).

Algorithm~\ref{alg:msf} shows an efficient implementation of \Boruvka's algorithm
based on shortcutting using pointer-jumping instead of contraction~\cite{zhou2017mst}.
The \Boruvka{} step is described in terms of the edgelist representation of a
graph. We note that in practice, it is often the case that the MSF is contained
in the first $2n$ or so lowest weight edges. This motivates running several
\Boruvka{} steps on prefixes of the edges sorted by weight, an optimization
which is also implemented by other experimental MSF
algorithms~\cite{osipov2009filter, shun2012brief, zhou2017mst}. Each run
computes a new set of edges that are now in the MSF which can be used to filter
the remaining edges that are now shortcut by the newly added edges. After
running a constant number of filter steps, we can run \Boruvka{} on the
remaining edges in the graph.

We now give a high-level description of the edgelist based \Boruvka{}
implementation and refer to~\cite{zhou2017mst} for a detailed explanation of
the code. Each vertex is initially its own parent, all vertices are active and
the initial forest is empty (Line 2). The
algorithm runs over a series of rounds. In each round, we initialize cells for
currently active vertices (Line 4). Next, we loop in parallel over all edges
and perform a priority-write with min based on the weight on both endpoints of
the edge (Lines 6 and 7). This writes the weight and index-id of a minimum-weight
edge incident to a vertex $v$ into $P[v]$. On Line 8, we mark the roots of the
forest (we break symmetry by marking the higher endpoint of an edge as the root
of that edge). On Line 9 we filter vertices that did not have any vertices join
their component and update the number of vertices that are still active. Next,
we update the forest, adding all edges which won on either of their endpoints in
$P$ (Line 10). We then use pointer-jumping to map every vertex to the id of
its new component root (Line 11) and relabel all edges based on the new ids of
each endpoint (Line 12). Finally, we filter any self-edges from the graph. We
note that our implementation uses indirection over both the vertices and edges
by maintaining a set of active vertices, of size $\codevar{active}$ and a set
of active edge-ids.
This helps improve performance in practice as we can allocate $P$ to have size
proportional to the number of active vertices in each round, and as we can
filter just the ids of the edges, instead of triples containing the two
endpoints and the weight of each edge.

\begin{tboxalg}{Strongly Connected Components} \label{alg:scc}
\small
\begin{algorithmic}[1]
\Procedure{SCC}{$G(V, E)$}
\State $\codevar{P} = \{V_{1}, \ldots, V_{\log n}\}  $ = \textproc{Partition}($V$)
\State $\codevar{L} = \{\infty, \ldots, \infty\}$, $\codevar{Done} = \{0, \ldots, 0\}$
\For{$i \leftarrow 1,\ldots, \log n$}
  \State $\codevar{Centers}$ = $\{v \in V_{i} \mid\ \codevar{Done}[i] == 0\}$
  \State $\codevar{InL} = \textproc{MarkReachable}(G^\mathsf{T}, L, \codevar{Centers})$
  \State $\codevar{OutL} = \textproc{MarkReachable}(G, L, \codevar{Centers})$
  \State Set $\codevar{Done}[i] = \codevar{true}, (i, \_\ ) \in \codevar{InL} \cap \codevar{OutL}$
  \State Set $L[i] = \min \{\codevar{label} \mid (i, \codevar{label}) \in \codevar{InL} \cap \codevar{OutL}\}$
  \State Set $L[i] = \min \{\codevar{label} \mid (i, \codevar{label}) \in \codevar{InL} \oplus \codevar{OutL}\ \text{and}\ !\codevar{Done}[i]\}$
\EndFor
\EndProcedure
\end{algorithmic}
\end{tboxalg}

\myparagraph{Strongly Connected Components}
Tarjan's algorithm is the textbook sequential algorithm for computing the
strongly connected components (SCCs) of a directed graph~\cite{CLRS}. As it uses depth-first
search, we currently do not know how to efficiently parallelize it~\cite{aggarwal1989parallel}.
The current theoretical state-of-the-art
for parallel SCC algorithms with polylogarithmic depth reduces the problem to computing
the transitive closure of the graph. This requires $\tilde{O}(n^3)$ work using
combinatorial algorithms~\cite{gazit1988improved}, which is
significantly higher than the $O(m + n)$ work done by sequential algorithms.
As the transitive-closure based approach performs a significant amount of work even for moderately sized graphs,
subsequent research on parallel SCC algorithms has focused on improving the
work while potentially sacrificing depth~\cite{fleischer2000identifying,
coppersmith2003divide, schudy2008finding, blelloch2016parallelism}.
Conceptually, these algorithms first pick a random pivot and use a
reachability-oracle to identify the SCC containing the pivot. They then remove
this SCC, which partitions the remaining graph into several disjoint pieces,
and recurse on the pieces.

In this paper, we present the first implementation of the SCC algorithm from Blelloch et
al.~\cite{blelloch2016parallelism}, shown in Algorithm~\ref{alg:scc}.
We refer the reader to Section 6.2 of~\cite{blelloch2016parallelism}
for proofs of correctness and its work and depth bounds. The algorithm
is similar in spirit to randomized quicksort. On Line 2 we randomly
permute the vertices and assign them to $\log n$ batches. On Line 3 we
set the initial label for all vertices as $\infty$ and mark all
vertices as not done. Now, we process the batches one at a time.  For
each batch, we compute $\codevar{Centers}$, which are the vertices in
the batch that are not yet done (Line 5). The next step calls
$\textproc{MarkReachable}$ from the centers on both $G$ and the
transposed graph, $G^\mathsf{T}$ (Lines 6--7).
$\textproc{MarkReachable}$ takes the set of centers and uses a
breadth-first search to compute the sets $\codevar{InL}$
($\codevar{OutL}$), which for a center $c \in \codevar{Centers}$
includes all $(v, c)$ pairs for vertices $v$ that $c$ can reach
through its in-edges (out-edges) that share the same label as $c$. On
Line 8, we mark all vertices that had a center visit them through both
the in-search and out-search as done---these vertices are captured by
some SCC. On Line 9, we deterministically set the label for vertices
that were finished in this round.  Finally, on Line 10, we set the
labels for vertices that were not finished in this round but were
visited by either the in or out-search.

Our implementation runs in in $O(m\log n)$ expected work and
$O(\mathsf{diam}(G)\log n)$ depth w.h.p. on the \pwmod{}-\mpram{}. One of
the challenges in implementing this SCC algorithm is how to compute
reachability information from multiple vertices simultaneously and how
to combine the information to (1) identify SCCs and (2) refine the
subproblems of visited vertices.
In our implementation, we explicitly store $\mathcal{R}_{F}$ and
$\mathcal{R}_{B}$, the forward and backward reachability sets for the set of
centers that are active in the current phase, $C_{A}$. The sets are represented as
hash tables that store tuples of vertices and center IDs, $(u, c_{i})$, representing a vertex $u$ in the
same subproblem as $c_{i}$ that is visited by a directed path from $c_{i}$.
We explain how to make the hash table technique practical in
Section~\ref{sec:techniques}.
The reachability sets are computed by running simultaneous breadth-first searches
from all active centers. In each round of the BFS, we apply \emap{} to
traverse all out-edges (or in-edges) of the current frontier. When we
visit an edge $(u,v)$ we try to add $u$'s center IDs to $v$. If $u$ succeeds in
adding any IDs, it test-and-set's a visited flag for
$v$, and returns it in the next frontier if the test-and-set
succeeded. Each BFS requires at most $O(\mathsf{diam}(G))$
rounds as each search adds the same labels on each round as it would
have had it run in isolation.


After computing $\mathcal{R}_{F}$ and $\mathcal{R}_{B}$, we
deterministically assign (with respect to the random permutation of
vertices generated at the start of the algorithm) vertices that we
visited in this phase a new label, which is either the label of a
refined subproblem or a unique label for the SCC they are contained
in. We first intersect the two tables and perform, for any tuple $(v,
c_{i})$ contained in the intersection, a priority-write with min on
the memory location corresponding to $v$'s SCC label with $c_{i}$ as
the label. Next, for all pairs $(v, c_{i})$ in $\mathcal{R}_{F} \oplus
\mathcal{R}_{B}$ we do a priority-write with min on $v$'s subproblem
label, which ensures that the highest priority search that visited $v$
sets its new subproblem.

We implemented an optimized search for the first phase, which just
runs two regular BFSs over the in-edges and out-edges from a single
pivot and stores the reachability information in bit-vectors instead
of hash-tables. It is well known that many directed real-world graphs
have a single massive strongly connected component, and so with
reasonable probability the first vertex in the permutation will find
this giant component~\cite{broder2000graph}. We also implemented a
`trimming' optimization that is reported in the
literature~\cite{mclendon2005finding, slota2014bfs}, which eliminates
trivial SCCs by removing any vertices that have zero in- or
out-degree.  We implement a procedure that recursively trims until no
zero in- or out-degree vertices remain, or until a maximum number of
rounds are reached.


\subsection{Covering Problems}\label{subsec:covering}
\begin{tboxalg}{Maximal Independent Set} \label{alg:mis}
\small
\begin{algorithmic}[1]
\State $\codevar{P} = \textproc{RandomPermutation}(\{0, \ldots, n-1\})$
\State $\codevar{Priority} = \{\textproc{NghsBefore}(v_{0}), \ldots, \textproc{NghsBefore}(v_{n-1})\}$

\Procedure{NghsBefore}{$v$} \algorithmicreturn{} $|\{u \in N(v) \mid \codevar{P}[u] < \codevar{P}[v]\}|$
\EndProcedure
\Procedure{MIS}{$G(V, E)$}
\State $\codevar{roots} = \{v \in V \mid \codevar{Priority}[v] = 0\}$
\State $\codevar{finished} = 0$, $I = \{ \}$
\While {$\codevar{finished} < |V|$}
  \State $I = I \cup \codevar{roots}$
  \State $\codevar{removed} = \{v \in V \mid v \in N(\codevar{roots} \text{ and } \codevar{Priority}[v] > 0\}$
  \State Set $\codevar{Priority}[v] = 0$ for all $v \in \codevar{removed}$
  \State $\codevar{finished} = \codevar{finished} + |\codevar{roots}| + |\codevar{removed}|$
  \State $\codevar{roots} = \textproc{edgeMap}(G, \codevar{removed}, \textproc{DecrementPriority})$
\EndWhile
\State \algorithmicreturn{} $I$
\EndProcedure
\end{algorithmic}
\end{tboxalg}

\myparagraph{Maximal Independent Set}
Maximal independent set (MIS) and maximal matching (MM) are easily solved in linear work
sequentially using greedy algorithms. Many efficient parallel maximal
independent set and matching algorithms have been developed over the
years~\cite{karp1984mis, luby1986mis, alon1986mis, israeli1986improved, blelloch2012greedy, Birn13}.
Blelloch et al. show that when the vertices
(or edges) are processed in a random order, the sequential greedy algorithms
for MIS and MM can be parallelized efficiently and give practical
algorithms~\cite{blelloch2012greedy}. Recently, Fischer and Noever
showed an improved depth bound for this algorithm~\cite{FischerN18}.

In this paper, we implement the rootset-based algorithm for MIS from
Blelloch et al.~\cite{blelloch2012greedy} which runs in $O(m)$ expected work and
$O(\log^2 n)$ depth w.h.p. on the \famod{}-\mpram{}. To the best of our
knowledge this is the first implementation of the rootset-based
algorithm; the implementations from~\cite{blelloch2012greedy} are
based on processing appropriately-sized prefixes of an order generated
by a random permutation $P$. Our implementation of the rootset-based
algorithm works on a priority-DAG defined by directing edges in the
graph from the higher-priority endpoint to the lower-priority
endpoint. On each round, we add all roots of the DAG into the
MIS, compute $N(\codevar{roots})$, the neighbors of the rootset that
are still active, and finally decrement the priorities of
$N(N(\codevar{roots}))$. As the vertices in $N(\codevar{roots})$ are
at arbitrary depths in the priority-DAG, we only decrement the
priority along an edge $(u, v)$, $u \in N(\codevar{roots})$ if $P[u] <
P[v]$. The algorithm runs in $O(m)$ work as we process each edge
once; the depth bound is $O(\log^2 n)$ as the priority-DAG has
$O(\log n)$ depth w.h.p.~\cite{FischerN18}, and each round takes
$O(\log n)$ depth. We were surprised that this implementation usually
outperforms the prefix-based implementation
from~\cite{blelloch2012greedy}, while also being simple to
implement.

The rootset-based MIS algorithm from Blelloch et al.~\cite{blelloch2012greedy}
is shown in Algorithm~\ref{alg:mis}. In MIS, we first randomly order the vertices
with a random permutation $P$ and compute the array
$\codevar{Priority}$, an array which for each vertex $v$ contains the
number of neighbors that have higher priority than $v$ by calling
$\textproc{NghsBefore}(v)$ according to $P$ (Lines 1--2). On Line 5 we
compute the initial rootset, $\codevar{roots}$, which is the set of
all vertices that have priority 0.  In each round, the algorithm adds
the roots to the independent set (Line 8), computes the set of removed
vertices, which are neighbors of the rootset that are still active
($\codevar{Priority}[v] > 0$). On Lines 10-11 we set the priority of
the removed vertices to 0 and update the number of finished vertices.
Finally, we compute the new rootset by decrementing the priority of
all edges $(u,v)$ where $u \in \codevar{removed}$ and $P[u] < P[v]$
using a fetch-and-add, and returning true for a neighbor $v$ if we
decrement its priority to 0.

\begin{tboxalg}{Maximal Matching} \label{alg:maxmatch}
\small
\begin{algorithmic}[1]
\Procedure{ParallelGreedyMM}{$P, \codevar{matched}$}
\State $\codevar{M} = \{ \}$
\While{$|P| > 0$}
  \State $\codevar{W} =$ edges in $P$ with no adjacent edges with higher priority
  \State Update $\codevar{matched}[u], \codevar{matched}[v]$ for all $(u,v) \in W$
  \State Filter edges incident to newly matched vertices from $P$
\EndWhile
\State \algorithmicreturn{} $M$
\EndProcedure
\medskip
\Procedure{MaximalMatching}{$G(V, E)$}
\State $\codevar{matched} = \{0, \ldots, 0\}, \codevar{M} = \{\}$
\While {$|E| > 0$}
  \State $\codevar{P}$ = Select a $1/d_{e}$-prefix of $E$
  \State $W = \textproc{ParallelGreedyMM}(P, \codevar{matched})$
  \State $E = E \setminus (W \cup N(W))$
  \State $M = M \cup W$
\EndWhile
\State \algorithmicreturn{} $M$
\EndProcedure
\end{algorithmic}
\end{tboxalg}

\myparagraph{Maximal Matching}
Our maximal matching implementation is based on the prefix-based
algorithm from~\cite{blelloch2012greedy} that takes $O(m)$ expected work and $O(\log^3m/\log\log m)$ depth w.h.p. on the \pwmod{}-\mpram{} (using the improved depth shown in~\cite{FischerN18}). We had to make several
modifications to run the algorithm on the large graphs in our
experiments. The original code from~\cite{blelloch2012greedy} uses an edgelist
representation, but we cannot directly use this implementation as uncompressing
all edges would require a prohibitive amount of memory for large graphs.
Instead, as in our MSF implementation, we simulate the prefix-based approach by
performing a constant number of \emph{filtering} steps. Each filter step packs out
$3n/2$ of the highest priority edges, randomly permutes them, and then
runs the edgelist based algorithm on the prefix.  After computing
the new set of edges that are added to the matching, we filter the remaining
graph and remove all edges that are incident to matched vertices. In practice, just
3--4 filtering steps are sufficient to remove essentially all edges in the
graph. The last step uncompresses any remaining edges into an edgelist and runs
the prefix-based algorithm. The filtering steps can be done within the work and depth bounds of the original algorithm.

The prefix-based maximal matching algorithm from Blelloch et
al.~\cite{blelloch2012greedy} is shown in Algorithm~\ref{alg:maxmatch}.
The algorithm first sets all vertices as unmatched, and initializes the
matching to empty (Line 9). The algorithm runs over a set of rounds; each
round selects a $1/d_{e}$-prefix of the edges ($d_e$ is the maximum number of neighboring edges an edge has), and runs the parallel greedy maximal
matching algorithm (Line 1) on it. The parallel greedy algorithm repeatedly
finds the set of edges that have the highest priority amongst all other edges
incident to either endpoint (Line 4), adds them to the matching (Line 5), and
filters the prefix based on the newly matched edges (Line 6). The edges matched
by the greedy algorithm are returned to the MaximalMatching procedure (Line
12). We then recompute $E$ by removing the matched edges, and the one-hop
neighborhood of the matched edges. Implementing this in practice can be done by
lazily deleting edges, and using doubling to select the next prefix.
We refer to~\cite{blelloch2012greedy} for the proof of the work and depth of
this algorithm. Our actual implementation does several filtering steps before executing Algorithm~\ref{alg:maxmatch} as described in Section~\ref{sec:algs}. This does not affect the work and depth bounds.

\begin{tboxalg}{LLF Graph Coloring} \label{alg:coloring}
\small
\begin{algorithmic}[1]
\State $\codevar{P} = \textproc{RandomPermutation}(\{1, \ldots, n-1\})$, $C = \{\infty, \ldots, \infty\}$
\State $\codevar{Priority}$ = priority based on log-degree, breaking ties using $\codevar{P}$
\Procedure{LLF}{$G(V, E)$}
\State $\codevar{roots} = \{v \in V | \codevar{Priority}[v] == 0\}$, $\codevar{finished} = 0$
\While {$\codevar{finished} < n$}
  \State $\textproc{AssignColors}(\codevar{roots})$
  \State $\codevar{finished} = \codevar{finished} + |\codevar{roots}|$
  \State $\codevar{roots} = \textproc{edgeMap}(G, \codevar{roots}, \textproc{DecrementPriority})$
\EndWhile
\State \algorithmicreturn{} $C$
\EndProcedure
\end{algorithmic}
\end{tboxalg}

\myparagraph{Graph Coloring}
As graph coloring is $\mathsf{NP}$-hard to solve optimally, algorithms like
greedy coloring, which guarantees a $(\Delta+1)$-coloring, are used instead in
practice, and often use much fewer than $(\Delta + 1)$ colors on real-world
graphs~\cite{welsh1967upper, hasenplaugh2014ordering}.
Jones and Plassmann (JP) parallelize the greedy algorithm using linear work,
but unfortunately adversarial inputs exist for the heuristics they consider
that may force the algorithm to run in $O(n)$ depth.
Hasenplaugh et al. introduce several heuristics that produce high-quality
colorings in practice and also achieve provably low-depth regardless of the
input graph. These include LLF (largest-log-degree-first), which processes
vertices ordered by the log of their degree and SLL (smallest-log-degree-last),
which processes vertices by removing all lowest log-degree vertices from the graph,
coloring the remaining graph, and finally coloring the removed vertices.
For LLF, they show that it runs in $O(m+n)$ work and $O(L \log
\Delta + \log n)$ depth, where $L=\min\{\sqrt{m}, \Delta\} + \log^{2}\Delta \log{n}/\log\log{n}$ in expectation.


In this paper, we implement a synchronous version of Jones-Plassmann
using the LLF heuristic in Ligra, which runs in $O(m + n)$ work and $O(L \log
\Delta + \log n)$ depth on the \famod{}-\mpram{}. The algorithm is implemented similarly
to our rootset-based algorithm for MIS. In each round, after coloring the roots
we use a fetch-and-add to decrement a count on our neighbors, and add
the neighbor as a root on the next round if the count is decremented
to 0.

Algorithm~\ref{alg:coloring} shows a synchronous implementation of the parallel LLF-Coloring
algorithm from~\cite{hasenplaugh2014ordering}.
On Lines 1 and 2 we compute a random permutation $P$, and compute
$\codevar{Priority}$ which for each vertex $v$ contains the number of neighbors
that have higher priority than $v$ based on comparing $\lceil \log d(v) \rceil$
and breaking ties using $P$. Initially, $\codevar{roots}$ is the set of all
vertices that have priority 0 (Line 4). In each round, we assign the lowest available color
the roots and update $\codevar{finished}$ (Line 6). We compute the next rootset by
decrementing the priority of all edges $(u,v)$ where $u \in \codevar{roots}$
using a fetch-and-add, and returning any neighbor whose priority is decremented
to 0 (Line 8).

\begin{tboxalg}{Approximate Set Cover} \label{alg:setcover}
\small
\begin{algorithmic}[1]
\Procedure{AboveThreshold}{$s$, $\codevar{d}$} \Return $\codevar{d} >= \lceil(1 + \epsilon)^{\max(b, 0)}\rceil$
\EndProcedure
\Procedure{SC}{$G = (S \cup E, A), \epsilon$}
\State $B$ = bucket sets based on degree and $\epsilon$, $\codevar{finished} = 0$
\State $C = \{ \}$
\While {$\codevar{finished} < n$}
  \State $(b, \codevar{sets}) = B.\textproc{NextBucket}()$ \Comment{Extract the highest bucket}
  \State Pack out neighbors of $\codevar{sets}$ that are covered
  \State $(S_{C}, S_{R})$ = split $\codevar{sets}$ based on new degrees
  \State Sets in $S_{C}$ try acquiring elements with a randomly chosen priority
  \State $A_{S}$ = $\{s \in S_{C} \mid s \text{ acquired enough neighbors}\}$
  \State $C = C \cup A_{S}$
  \State Reset neighbors of $S_{C}$ that are not covered
  \State $B.\textproc{UpdateBuckets}((S_{C} \setminus A_{S}) \cup S_{R})$
 \EndWhile
\State \algorithmicreturn{} $C$
\EndProcedure
\end{algorithmic}
\end{tboxalg}

\myparagraph{Approximate Set Cover}
The set cover problem can be modeled by a bipartite graph where sets and
elements are vertices, with an edge between a set an element if and only if the
set covers that element. Like graph coloring, the set cover problem is
$\mathsf{NP}$-hard to solve optimally, and a sequential greedy algorithm
computes an $H_{n}$-approximation in $O(m)$ time for unweighted sets, and
$O(m\log m)$ time for weighted sets, where $H_n=\sum_{k=1}^{n}1/k$ and $m$ is
the sum of the sizes of the sets (or the number of edges in the graph). There
has been significant work on finding work-efficient parallel algorithms that
achieves an $H_{n}$-approximation~\cite{Berger1994,Rajagopalan1999,blelloch11manis,blelloch12setcover,Kumar2015}.

Algorithm~\ref{alg:setcover} shows pseudocode for the Blelloch et al.
algorithm~\cite{blelloch11manis} which runs in $O(m)$ work and $O(\log^3 n)$
depth on the \pwmod{}-\mpram{}. We refer to~\cite{dhulipala2017julienne} for a detailed
explanation of the code, and give a high-level description of the algorithm
here. The algorithm first buckets the sets based on their degree, placing a set
covering $D$ elements into $\lfloor \log_{1+\epsilon} D \rfloor$'th bucket
(Line 2). We process the buckets in decreasing order. In each round, we extract
the highest bucket (Line 5) and pack out their adjacency lists to remove neighbors
that may have been covered by prior rounds (Line 6). We then split the sets into
$S_{C}$, sets that continue in the round, and $S_{R}$, sets that should be rebucketed based
on whether the set still contains enough uncovered neighbors (Line 8). Next, we
implement one step of MaNIS~\cite{blelloch11manis}, which subselects a set of
sets from $S_{C}$ that have little overlap. We select a random priority for
each set in $S_{C}$ and try to acquire all neighbors of the set (Line 9) by doing
a priority-write with min using the randomly chosen priority. On Line 10, we
compute $A_{S}$, the sets that won on more than $\lceil(1 + \epsilon)^{\max(b-1, 0)}\rceil$
of their neighbors, and add them to the cover (Line 11). Next, we reset memory
locations for elements that were written-to by a set, but not acquired (Line
12).  Finally we reinsert sets that did not win on enough neighbors ($S_{C}
\setminus A_{S})$, or lost enough of their size ($S_{R}$) back into the bucket
structure.

Our implementation of approximate set cover in this paper is based on the
implementation from Julienne~\cite{dhulipala2017julienne}. The main change
we made in this paper is to ensure that we correctly set random priorities
on each round of the algorithm. Both the implementation in Julienne as well as
an earlier implementation of the algorithm~\cite{blelloch12setcover} use vertex
IDs instead of picking random priorities for all sets that are active on a given
round. This can cause very few vertices to be added on each round on meshes and
other graphs with a large amount of symmetry. Instead, in our implementation,
for $\mathcal{A}_{S}$, the active sets on a round, we generate a random
permutation of $[0, \ldots, |\mathcal{A}_{S}|-1]$ and write these values into a
pre-allocated dense array with size proportional to the number of sets. We
give experimental details regarding this change in Section~\ref{sec:exps}.

\subsection{Substructure Problems}\label{subsec:substructure}
\begin{tboxalg}{$k$-core} \label{alg:kcore}
\small
\begin{algorithmic}[1]
\Procedure{Coreness}{$G(V, E)$}
\State $D = \{\emph{deg}(v_{0}), \ldots, \emph{deg}(v_{n-1})\}$
\State $B$ = bucket vertices based on D, $\codevar{finished} = 0$
\While {$\codevar{finished} < n$}
  \State $(k, ids) = B.\textproc{NextBucket}()$
  \State $\codevar{finished} = \codevar{finished} + |ids|$
  \State $\codevar{moved} = \textproc{Histogram}(G, ids, \textproc{DecrementCoreness})$
  \State $B.\textproc{UpdateBuckets}(\codevar{moved})$
\EndWhile
\State \algorithmicreturn{} $D$
\EndProcedure
\end{algorithmic}
\end{tboxalg}

\myparagraph{$k$-core}
 $k$-cores were
defined independently by Seidman~\cite{seidman83network}, and by
Matula and Beck~\cite{matula83smallest} who also gave a linear-time
algorithm for computing the \emph{coreness} value of all vertices,
i.e. the maximum $k$-core a vertex participates in. Anderson and Mayr
showed that $k$-core (and therefore coreness) is in
$\textsf{NC}$ for $k \leq 2$, but is $\textsf{P}$-complete for $k \geq
3$ \cite{anderson84pcomplete}. The Matula and Beck algorithm is simple
and practical---it first bucket-sorts vertices by their degree, and
then repeatedly deletes the minimum-degree vertex. The affected
neighbors are moved to a new bucket corresponding to their induced
degree. As each edge in each direction and vertex is processed exactly
once, the algorithm runs in $O(m + n)$
work. In~\cite{dhulipala2017julienne}, the authors give a parallel
algorithm based on bucketing that runs in $O(m + n)$ expected work,
and $\rho \log n$ depth w.h.p. $\rho$ is the peeling-complexity of the
graph, defined as the number of rounds to peel the graph to an empty
graph where each peeling step removes all minimum degree
vertices.

Our implementation of $k$-core in this paper is based on the implementation
from Julienne~\cite{dhulipala2017julienne}. One of the challenges to implementing
the peeling algorithm for $k$-core is efficiently computing the number of edges
removed from each vertex that remains in the graph. A simple approach is to
just fetch-and-add a counter per vertex, and update the bucket of the vertex based
on this counter, however this incurs significant contention on real-world graphs with
vertices with large degree. In order to make this step faster in
practice, we implemented a work-efficient histogram which computes the
number of edges removed from remaining vertices while incurring very little
contention. We describe our histogram implementation in Section~\ref{sec:techniques}.

Algorithm~\ref{alg:kcore} shows pseudocode for the work-efficient $k$-core
algorithm from Julienne~\cite{dhulipala2017julienne} which computes the
coreness values of all vertices.  The algorithm initializes the initial
coreness values to the degree of each vertex (Line 2), and inserts the vertices
into a bucketing data-structure based on their degree (Line 3). In each round,
while all of the vertices have not yet been processed the algorithm removes the
vertices in the minimum bucket (Line 5), computes the number of edges removed
from each neighbor using a histogram (Line 7) and finally updates the buckets
of affected neighbors (Line 8). We return the array $D$, which contains the
coreness values of each vertex at the end of the algorithm.


\begin{tboxalg}{Approximate Densest Subgraph} \label{alg:densest-subgraph}
\small
\begin{algorithmic}[1]
\State $D = \{\emph{deg}(v)\ |\ v \in V\}$\label{ds:induceddegs}

\Procedure{Cond}{$v$} \algorithmicreturn{} $1$
\EndProcedure
\Procedure{UpdateDecrement}{$s$, $d$}
\State $\textproc{\fa{}}(\&D[d], -1)$
\State \algorithmicreturn{} 0
\EndProcedure

\Procedure{ApproximateDensestSubgraph}{$G(V, E)$}
\State $S = V,\ S_{\max} = \emptyset$ \label{ds:init}
\While{$S \neq \emptyset$}\label{ds:whilestart}
  \State $R = \{v \in S\ |\ D[v] < 2(1+\epsilon)\rho(S)$\}\label{ds:computeR} \Comment{$\rho(S) = \frac{|E(G[S])|}{|S|}$, $G[S]$ is the induced subgraph on $S$}
  \State $\emap{}(G, R, \textproc{UpdateDecrement}, \textproc{Cond})$\label{ds:emap}
  \State $D = D \setminus R$\label{ds:removeR}
  \If{$\rho(D) > \rho(D_{\max})$} 
    \State $D_{\max} = D$
  \EndIf
\EndWhile\label{ds:whileend}
\State \algorithmicreturn{} $\codevar{D}_{\max}$
\EndProcedure
\end{algorithmic}
\end{tboxalg}

\myparagraph{Approximate Densest Subgraph}
The densest subgraph problem is to find a subgraph of an
undirected graph with the highest density (the density of a subgraph
is the number of edges in the subgraph divided by the number of
vertices). The problem is a classic graph optimization problem that
admits exact polynomial-time solutions using either a reduction to
flow~\cite{goldberg84densest} or
LP-rounding~\cite{charikar00densesubgraph}. In his paper, Charikar
also gives a simple $O(m+n)$ work 2-approximation algorithm based on computing a
degeneracy ordering of the graph, and taking the maximum density
subgraph over all suffixes of the degeneracy order\footnote{We note
that the $2$-approximation be work-efficiently solved in the same
depth as our $k$-core algorithm by augmenting the $k$-core algorithm to return
the order in which vertices are peeled. Computing the maximum density
subgraph over suffixes of the degeneracy order can be done using scan.}.
The problem has also received attention in parallel models of
computation~\cite{bahmani2012densest, bahmani2014efficient}.
Bahmani et al. give a $(2+\epsilon)$-approximation running in $O(\log_{1+\epsilon} n)$
rounds of MapReduce~\cite{bahmani2012densest}. Subsequently, Bahmani et
al.~\cite{bahmani2014efficient} showed that a $(1+\epsilon)$ can be
found in $O(\log n / \epsilon^2)$ rounds of MapReduce by using
the multiplicative-weights approach on the dual of the natural LP for
densest subgraph. To the best our knowledge, it is currently open
whether the densest subgraph problem can be exactly solved in $\NC{}$.

In this paper, we implement the elegant $(2+\epsilon)$-approximation
algorithm of Bahmani et al. (Algorithm~\ref{alg:densest-subgraph}).
Our implementation of the algorithm runs in $O(m+n)$ work and
$O(\log_{1+\epsilon} n\log n)$ depth. The algorithm starts with a
candidate densest subgraph, $S$, consisting of all vertices, and
an empty densest subgraph $S_{\max}$ (Line~\ref{ds:init}). It also
maintains an array with the induced degree of each vertex in $S$,
which is initially just its degree in $G$ (Line~\ref{ds:induceddegs}).
The main loop iteratively peels vertices with degree below the density
threshold in the current candidate subgraph
(Lines~\ref{ds:whilestart}--\ref{ds:whileend}).
Specifically, it finds all vertices with induced degree less than
$2(1+\epsilon) \rho(S)$ (Line~\ref{ds:computeR}), calls \emap{}, which
updates the induced degrees array (Line~\ref{ds:emap}) and finally
removes them from $S$ (Line~\ref{ds:removeR}). If the density of the
updated subgraph $S$ is greater than the density of $S_{\max}$, the
algorithm updates $S_{\max}$ to be $S$.

Bahmani et al. show that this algorithm removes a constant factor of
the vertices in each round. However, they do not consider the work of
the algorithm in the MapReduce model. We briefly sketch how the
algorithm can be implemented in $O(m+n)$ work and
$O(\log_{1+\epsilon} n\log n)$ depth. Instead of computing the density
of the current subgraph by scanning all edges, we maintain it
explicitly in an array, $D$ (Line~\ref{ds:induceddegs}), and update it
as vertices are removed from $S$. Each round of the algorithm does
work proportional to vertices in $S$ to compute $R$
(Line~\ref{ds:computeR}) but since $S$ decreases by a constant factor
in each round the work of these steps is $O(n)$ over all rounds.
Computing the new density can be done by computing the number of edges
between $R$ and $S$, which only requires scanning edges incident to
vertices in $R$ using \emap{} (Line~\ref{ds:emap}).  Therefore, the
edges incident to a vertex are scanned exactly once, in the round when
it is included in $R$, and so the algorithm performs $O(m+n)$ work.
The depth is $O(\log_{1+\epsilon} n\log n)$ since there are
$O(\log_{1+\epsilon} n)$ rounds each of which perform a filter and
\emap{} which both run in $O(\log n)$ depth. In practice, we found
that since a large number of vertices are removed in each round, using
fetch-and-add can cause contention, especially on graphs containing
vertices with high degrees. Instead, our implementation uses a
work-efficient histogram procedure (see Section~\ref{sec:techniques})
which updates the degrees while incurring very little contention.

\myparagraph{Triangle Counting}
Triangle counting has received significant recent attention due to its
numerous applications in Web and social network analysis. There have
been dozens of papers on sequential triangle
counting~\cite{Itai1977,AYZ97,Schank2005,Schank2007,Latapy2008,OrtmannB14,PS2014}. The
fastest algorithms rely on matrix multiplication and run in either
$O(n^\omega)$ or $O(m^{2\omega/(1+\omega)})$ work, where $\omega$ is
the best matrix multiplication exponent~\cite{Itai1977,AYZ97}. The
fastest algorithm that does not rely matrix multiplication requires
$O(m^{3/2})$ work~\cite{Schank2005,Schank2007,Latapy2008}, which also
turns out to be much more practical. Parallel algorithms with
$O(m^{3/2})$ work have been
designed~\cite{shun2015multicore,aberger2017empty,low2010graphlab},
with Shun and Tangwongsan~\cite{shun2015multicore} showing an algorithm that requires $O(\log n)$ depth
on the \mpram{}.\footnote{The algorithm in~\cite{shun2015multicore} was described in the Parallel Cache Oblivious model, with a depth of $O(\log^{3/2}n)$.}

The implementation from~\cite{shun2015multicore} parallelizes Latapy's
\emph{compact-forward} algorithm, which creates a directed graph $DG$
where an edge $(u,v) \in E$ is kept in $DG$ iff $\emph{deg}(u) < \emph{deg}(v)$.
Although triangle counting can be done directly on the undirected graph in the
same work and depth asymptotically, directing the edges helps reduce work, and
ensures that every triangle is counted exactly once.

In this paper we implement the triangle counting algorithm described
in~\cite{shun2015multicore}.
We had to make several significant changes to the implementation
in order to run efficiently on large compressed graphs. First, we parallelized
the creation of the directed graph; this step creates a directed graph encoded
in the parallel-byte format in $O(m)$ work and $O(\log n)$ depth. We also parallelized
the merge-based intersection algorithm to make it work in the parallel-byte format.
We give more details on these techniques in
Section~\ref{sec:techniques}.

\subsection{Eigenvector Problems}\label{subsec:eigenvector}
\myparagraph{PageRank}
PageRank is a centrality algorithm first used at Google to rank
webpages~\cite{brin1998pagerank}. The algorithm takes a graph
$G=(V,E)$, a damping factor $0 \leq \gamma \leq 1$ and a constant
$\epsilon$ which controls convergence. Initially, the PageRank of each
vertex is $1/n$. On each iteration, the algorithm updates the
PageRanks of the vertices using the following equation:
\begin{equation*}
  P_{v} = \frac{1-\gamma}{n} + \gamma \sum_{u \in N^{-}(v)}
  \frac{P_{u}}{\emph{deg}^{+}(u)}
\end{equation*}
This update step can be implemented using a single sparse-matrix
vector multiplication call (implementable using, for example,
\emap{}).  The algorithm implemented in this paper follows the
implementation of PageRank described in
Ligra~\cite{shun2012ligra}. We note that many PageRank implementations
in the wild actually implement an algorithm called PageRank-Delta,
which modified PageRank by only activating a vertex if its PageRank
value has changed sufficiently. However, we are not aware of any
bounds on the work and depth of this algorithm, and therefore
chose to implement the classic PageRank.

The main modification we made to the implementation from Ligra was to
implement the dense iterations of the algorithm using a reduction
primitive, which can be carried out over the incoming neighbors of a
vertex in parallel, without needing a fetch-and-add instruction. Each
iteration of our implementation requires $O(m+n)$ work and $O(\log n)$
depth.

\section{Implementations and Techniques}\label{sec:techniques}

\algblock{ParFor}{EndParFor}
\algnewcommand\algorithmicparfor{\textbf{parfor}}
\algnewcommand\algorithmicpardo{\textbf{do}}
\algnewcommand\algorithmicendparfor{}
\algrenewtext{ParFor}[1]{\algorithmicparfor\ #1\ \algorithmicpardo}
\algrenewtext{EndParFor}{\algorithmicendparfor}
\algtext*{EndParFor}{}

In this section, we introduce several general implementation techniques and
optimizations that we use in our algorithms. The techniques include a
fast histogram implementation useful for reducing contention in the
$k$-core algorithm, a cache-friendly sparse \emap{} implementation
that we call \emapblock{}, and compression techniques used to
efficiently parallelize algorithms on massive graphs.

\subsection{A Work-efficient Histogram
  Implementation}\label{subsec:historgram}
Our initial implementation of the peeling-based algorithm for $k$-core
algorithm suffered from poor performance due to a large amount of contention
incurred by fetch-and-adds on high-degree vertices. This occurs as many social-networks
and web-graphs have large maximum degree, but relatively small degeneracy,
or largest non-empty core (labeled $k_{max}$ in Table~\ref{table:sizes}).
For these graphs, we observed that many early rounds, which process vertices
with low coreness perform a large number of fetch-and-adds on memory locations
corresponding to high-degree vertices, resulting in high
contention~\cite{shun13reducing}.  To reduce contention, we designed a work-efficient
histogram implementation that can perform this step while only incurring
$O(\log n)$ contention w.h.p. The \defn{Histogram} primitive takes a sequence of
$(\bf{K}, \bf{T})$ pairs, and an associative and commutative operator $R:
{\bf T} \times {\bf T} \rightarrow {\bf T}$ and computes a sequence of
$(\bf{K}, \bf{T})$ pairs, where each key $k$ only appears once, and its
associated value $t$ is the sum of all values associated with keys $k$ in the
input, combined with respect to $R$.

A useful example of histogram to consider is summing for each $v \in
N(F)$ for a \vset{} $F$, the number of edges $(u,v)$ where $u \in F$
(i.e., the number of incoming neighbors from the frontier).  This
operation can be implemented by running histogram on a sequence where
each $v \in N(F)$ appears once per $(u,v)$ edge as a tuple $(v, 1)$
using the operator $+$. One theoretically efficient implementation of
histogram is to simply semisort the pairs using the work-efficient
semisort algorithm from~\cite{gu15semisort}. The semisort places pairs
from the sequence into a set of \emph{heavy} and \emph{light} buckets,
where heavy buckets contain a single key that appears many times in
the input sequence, and light buckets contain at most $O(\log^2 n)$
distinct keys $(k, v)$ keys, each of which appear at most $O(\log n)$
times w.h.p. (heavy and light keys are determined by sampling). We
compute the reduced value for heavy buckets using a standard parallel
reduction. For each light bucket, we allocate a hash table, and hash
the keys in the bucket in parallel to the table, combining multiple
values for the same key using $R$. As each key appears at most $O(\log
n)$ times w.h.p, we incur at most $O(\log n$) contention w.h.p. The
output sequence can be computed by compacting the light tables and heavy arrays.

While the semisort implementation is theoretically efficient, it
requires a likely cache miss for each key when inserting into the
appropriate hash table.  To improve cache performance in this step, we
implemented a work-efficient algorithm with $O(n^{\epsilon})$ depth
based on radix sort. Our implementation is based on the parallel radix
sort from PBBS~\cite{shun2012brief}. As in the semisort, we first
sample keys from the sequence and determine the set of heavy-keys.
Instead of directly moving the elements into light and heavy buckets,
we break up the input sequence into $O(n^{1-\epsilon})$ blocks, each
of size $O(n^{\epsilon})$, and sequentially sort the keys within a
block into light and heavy buckets.  Within the blocks, we reduce all
heavy keys into a single value and compute an array of size
$O(n^{\epsilon})$ which holds the starting offset of each bucket
within the block.  Next, we perform a segmented-scan~\cite{Blelloch93}
over the arrays of the $O(n^{1-\epsilon})$ blocks to compute the sizes
of the light buckets, and the reduced values for the heavy-buckets,
which only contain a single key. Finally, we allocate tables for the
light buckets, hash the light keys in parallel over the blocks and
compact the light tables and heavy keys into the output array. Each
step runs in $O(n)$ work and $O(n^{\epsilon})$ depth.  Compared to the
original semisort implementation, this version incurs fewer cache
misses because the light keys per block are already sorted and
consecutive keys likely go to the same hash table, which fits in
cache.  We compared our times in the histogram-based version of
$k$-core and the fetch-and-add-based version of $k$-core and saw
between a 1.1--3.1x improvement from using the histogram.

\begin{algorithm}[!t]
\caption{\emapblock{}}\label{alg:emapblock}
\small
\begin{algorithmic}[1]
\Procedure{\emapblock{}}{$G, U, F$}
\State $\codevar{O}$ = Prefix sums of degrees of $u \in U$
\State $d_{U} = \sum_{u \in U} \emph{deg}(u)$
\State $\codevar{nblocks} = \lceil d_{U} / \codevar{bsize} \rceil$
\State $\codevar{B}$ = Result of binary search for $\codevar{nblocks}$ indices into $\codevar{O}$
\State $\codevar{I}$ = Intermediate array of size $\sum_{u \in U} \emph{deg}(u)$
\State $\codevar{A}$ = Intermediate array of size $\codevar{nblocks}$
\ParFor {$i \in \codevar{B}$}
  \State Process work in $B[i]$ and pack live neighbors into $I[i\codevar{bsize}]$
  \State $\codevar{A}[i]$ = Number of live neighbors
\EndParFor
\State $\codevar{R}$ = Prefix sum $\codevar{A}$ and compact $\codevar{I}$
\State \algorithmicreturn{} $R$
\EndProcedure
\end{algorithmic}
\end{algorithm}

\subsection{\emapblock{}}\label{subsec:emapblocked}
One of the core primitives used by our
algorithms is \emap{} (described in Section~\ref{sec:prelims}). The
push-based version of \emap{}, \emapsparse{}, takes a frontier $U$ and
iterates over all $(u,v)$ edges incident to it. It applies an update
function on each edge that returns a boolean indicating whether or not
the neighbor should be included in the next frontier. The standard
implementation of \emapsparse{} first computes prefix-sums of
$\emph{deg}(u), u \in U$ to compute offsets, allocates an array of
size $\sum_{u \in U} \emph{deg(u)}$, and iterates over all $u \in U$
in parallel, writing the ID of the neighbor to the array if the update
function $F$ returns \emph{true}, and $\bot$ otherwise. It then filters
out the $\bot$ values in the array to produce the output \vset{}.

In real-world graphs, $|N(U)|$, the number of unique neighbors
incident to the current frontier is often much smaller than $\sum_{u
  \in U} \emph{deg(u)}$.  However, \emapsparse{} will always perform
$\sum_{u \in U} \emph{deg}(u)$ writes and incur a proportional number
of cache misses, despite the size of the output being at most
$|N(U)|$. More precisely, the size of the output is at most $LN(U)
\leq |N(U)|$, where $LN(U)$ is the number of \emph{live neighbors} of
$U$, where a live neighbor is a neighbor of the current frontier for
which $F$ returns \emph{true}.  To reduce the number
of cache misses we incur in the push-based traversal, we implemented a
new version of \emapsparse{} that performs at most $LN(U)$ writes that
we call \emapblock{}.  The idea behind \emapblock{} is to logically
break the edges incident to the current frontier up into a set of
blocks, and iterate over the blocks sequentially, packing live
neighbors, compactly for each block. We then simply prefix-sum the
number of live neighbors per-block, and compact the block outputs into
the output array.

We now describe a theoretically efficient implementation of
\emapblock{} (Algorithm~\ref{alg:emapblock}). As in \emapsparse{}, we
first compute an array of offsets $\codevar{O}$ (Line 1) by prefix
summing the degrees of $u \in U$. We process the edges incident to
this frontier in blocks of size $\codevar{bsize}$. As we cannot afford
to explicitly write out the edges incident to the current frontier to
block them, we instead logically assign the edges to blocks. Each
block searches for a range of vertices to process with
$\codevar{bsize}$ edges; the $i$'th block binary searches the offsets
array to find the vertex incident to the start of the $(i \cdot
\codevar{bsize})$'th edge, storing the result into $B[i]$ (Lines
4--5).  The vertices that block $i$ must process are therefore between
$B[i]$ and $B[i+1]$. We note that multiple blocks can be assigned to
process the edges incident to a high-degree vertex.  Next, we allocate
an intermediate array $I$ of size $d_{U}$ (Line 6), but do not
initialize the memory, and an array $A$ that stores the number of live
neighbors found by each block (Line 7). Next, we process the blocks in
parallel by sequentially applying $F$ to each edge in
the block and compactly writing any live neighbors to $I[i \cdot
  \codevar{bsize}]$ (Line 9), and write the number of live neighbors
to $A[i]$ (Line 10). Finally, we do a prefix sum on $A$, which gives
offsets into an array of size proportional to the number of live
neighbors, and copy the live neighbors in parallel to $R$, the output
array (Line 11).

We found that this optimization helps the most in algorithms where there is a
significant imbalance between the size of the output of each \emap{}, and
$\sum_{u \in U} \emph{deg}(u)$. For example, in weighted BFS, relatively few of the
edges actually relax a neighboring vertex, and so the size of the output, which
contains vertices that should be moved to a new bucket, is usually much smaller
than the total number of edges incident to the frontier. In this case, we
observed as much as a 1.8x improvement in running time by switching from
\emapsparse{} to \emapblock{}.

\subsection{Techniques for overlapping
  searches}\label{subsec:overlappingsearch}
In this section, we describe how we compute and update the reachability labels
for vertices that are visited in a phase of our SCC algorithm. Recall
that each phase performs a graph traversal from the set of active
centers on this round, $C_{A}$, and computes for
each center $c$, all vertices in the weakly-connected component for the
subproblem of $c$ that can be reached by a directed path from it. We
store this reachability information as a set of
$(u, c_{i})$ pairs in a hash-table, which represent the fact that $u$ can
be reached by a directed path from $c_{i}$. A phase performs two graph
traversals from the centers to compute $\mathcal{R}_{F}$ and $\mathcal{R}_{B}$,
the out-reachability set and in-reachability sets respectively. Each traversal
allocates an initial hash table and runs rounds of \emap{} until no new label
information is added to the table.

The main challenge in implementing one round in the traversal is (1) ensuring
that the table has sufficient space to store all pairs that will be added this
round, and (2) efficiently iterating over all of the pairs associated with a
vertex. We implement (1) by performing a parallel reduce to sum over vertices $u \in F$, the
current frontier, the number of neighbors $v$ in the same subproblem, multiplied by
the number of distinct labels currently assigned to $u$. This
upper-bounds the number of distinct labels that could be added this round, and although we may
overestimate the number of actual additions, we will never run out of space in the table.
We update the number of elements currently in the table during concurrent
insertions by storing a per-processor count which gets incremented whenever the
processor performs a successful insertion. The counts are then summed together
at the end of a round and used to update the count of the number of
elements in the table.

One simple implementation of (2) is to simply allocate $O(\log n)$
space for every vertex, as the maximum number of centers that visit
any vertex during a phase is at most $O(\log n)$ w.h.p. However, this
will waste a significant amount of space, as most vertices are visited
just a few times.  Instead, our implementation stores $(u, c)$ pairs
in the table for visited vertices $u$, and computes hashes based only
on the ID of $u$. As each vertex is only expected to be visited a
constant number of times during a phase, the expected probe length is
still a constant. Storing the pairs for a vertex in the same
probe-sequence is helpful for two reasons. First, we may incur fewer
cache misses than if we had hashed the pairs based on both entries, as
multiple pairs for a vertex can fit in the same cache line. Second,
storing the pairs for a vertex along the same probe sequence makes it
extremely easy to find all pairs associated with a vertex $u$, as we
simply perform linear-probing, reporting all pairs that have $u$ as
their key until we hit an empty cell.  Our experiments show that this
technique is practical, and we believe that it may have applications
in similar algorithms, such as computing least-element lists or FRT
trees in parallel~\cite{blelloch2016parallelism,blelloch2016frt}.

\subsection{Primitives on Compressed
  Graphs}\label{subsec:compressedprimitives}
Many of our algorithms are concisely expressed using fundamental primitives
such as map, map-reduce, filter, pack, and intersection. To run our
algorithms without any modifications on compressed graphs, we wrote new
implementations of these primitives using using the parallel-byte format from Ligra+,
some of which required some new techniques in order to be theoretically efficient.
We first review the byte and parallel-byte formats
from~\cite{shun2015ligraplus}. In byte coding, we store a vertex's neighbor
list by difference encoding consecutive vertices, with the first vertex
difference encoded with respect to the source. Decoding is done by sequentially
uncompressing each difference, and
summing the differences into a running sum which gives the ID of the next
neighbor. As this process is sequential, graph algorithms using the byte format
that map over the neighbors of a vertex will
require $O(\Delta)$ depth. The parallel-byte format from Ligra+ breaks the
neighbors of a high-degree vertex into blocks, where each block
contains a constant number of neighbors. Each block is difference encoded with
respect to the source. As each block can have a different size, it also stores
offsets that point to the start of each block. The format stores the blocks in
a neighbor list $L$ in sorted order.

We now describe efficient implementations of primitives used by our algorithms.
All descriptions are given for neighbor lists coded in the parallel-byte
format.  The \defn{Map} primitive takes as input neighbor list $L$, and a map
function $F$, and applies $F$ to each ID in $L$. This can be implemented with a
parallel-for loop across the blocks, where each iteration decodes
its block sequentially.
Our implementation of map runs in $O(|L|)$ work and $O(\log n)$ depth.
\defn{Map-Reduce} takes as input a neighbor list $L$, a
map function $F : {\bf vtx} \rightarrow {\bf T}$ and a binary associative
function $R$ and returns the sum of the mapped elements with respect to $R$.  We perform
map-reduce similarly by first mapping over the blocks, then sequentially reducing over
the mapped values in each block. We store the accumulated value on the stack
or in an allocated array if the number of blocks is large enough. Finally, we
reduce the accumulated values using $R$ to compute the output.
Our implementation of map-reduce runs in $O(|L|)$ work and $O(\log n)$ depth.

\defn{Filter} takes as input a neighbor list $L$, a
predicate $P$, and an array $T$ into which the vertices
satisfying $P$ are written, in the same order as in $L$. Our implementation of
filter also takes as input an array $S$, which is an array of size
$\emph{deg}(v)$ space for lists $L$ larger than a constant threshold, and
null otherwise. In the case where $L$ is large, we implement the filter by
first decoding $L$ into $S$ in parallel; each block in $L$ has an offset
into $S$ as every block except possibly the last block contains the same number of vertex IDs.
We then filter $S$ into the output array $T$. In the case where $L$ is small we
just run the filter sequentially.
Our implementation of filter runs in $O(|L|)$ work and $O(\log n)$ depth.
\defn{Pack} takes as input a neighbor list
$L$ and a predicate $P$ function, and packs $L$, keeping only vertex IDs
that satisfied $P$. Our implementation of pack takes as input an array $S$,
which an array of size $2*\emph{deg}(v)$ for lists larger than a constant threshold, and null otherwise. In the
case where $L$ is large, we first decode $L$ in parallel into the first
$\emph{deg}(v)$ cells of $S$. Next, we filter these vertices into the second
$\emph{deg}(v)$  cells of $S$, and compute the new length of $L$. Finally, we
recompress the blocks in parallel by first computing the compressed size of each new
block. We prefix-sum the sizes to calculate offsets into the array and finally
compress the new blocks by writing each block starting at its offset. When $L$ is
small we just pack $L$ sequentially. We make
use of the pack and filter primitives in our implementations of maximal
matching, minimum spanning forest, and triangle counting.
Our implementation of pack runs in $O(|L|)$ work and $O(\log n)$ depth.

The \defn{Intersection} primitive takes as input two neighbor lists
$L_{a}$ and $L_{b}$ and computes the size of the intersection of $L_{a}$ and
$L_{b}$ ($|L_{a}| \le |L_{b}|$). We implement an algorithm similar to the optimal
parallel intersection algorithm for sorted lists. As the blocks are compressed,
our implementation works on the first element of each block, which can be quickly
decoded. We refer to these elements as block starts. If the number of blocks in
both lists sum to less than a constant, we intersect them sequentially.
Otherwise, we take the start $v_s$ of the middle block in $L_{a}$, and binary
search over the starts of $L_{b}$ to find the first block whose start is less than or equal to $v_s$. Note that as the closest value less than or equal to $v_s$
could be in the middle of the block, the subproblems we generate
must to consider elements in the two adjoining blocks of each list, which adds
an extra constant factor of work in the base case. Our implementation of
intersection runs in $O(|L_{a}|\log(1 + |L_{b}|/|L_{a}|))$ work and $O(\log
n)$ depth.

\section{Experiments}\label{sec:exps}

In this section, we describe our experimental results on a set of
real-world graphs and also discuss related experimental work.
Tables~\ref{table:small-times} and ~\ref{table:large-times} show the running
times for our implementations on our graph inputs. For compressed
graphs, we use
the compression schemes from Ligra+~\cite{shun2015ligraplus}, which
we extended to ensure theoretical efficiency.
We describe these
modifications and also other statistics about our algorithms (e.g.,
number of colors used, number of SCCs, etc.) in
section~\ref{sec:graphstats}.

\subsection{Experimental Setup and Graph Inputs}
\myparagraph{Experimental Setup} We run all of our experiments on a 72-core
Dell PowerEdge R930 (with two-way hyper-threading) with $4\times 2.4\mbox{GHz}$
Intel 18-core E7-8867 v4 Xeon processors (with a 4800MHz bus and 45MB L3 cache)
and 1\mbox{TB} of main memory. Our programs use Cilk Plus to express
parallelism and are compiled with the \texttt{g++} compiler (version 5.4.1)
with the \texttt{-O3} flag. By using Cilk's work-stealing scheduler we are able obtain an expected running time of $W/P + O(D)$ for an algorithm with $W$ work and $D$ depth on $P$ processors~\cite{Blumofe1999}. For the parallel experiments, we use the command \texttt{numactl -i all} to balance the memory allocations across the sockets.
All of the speedup numbers we report are the running times of our
parallel implementation on 72-cores with hyper-threading over the
running time of the implementation on a single thread.

\begin{table}[!t]\footnotesize
\centering
\centering
\begin{tabular}[!t]{l|r|r|r|r|r}
\toprule
{Graph Dataset} & Num. Vertices & Num. Edges & $\mathsf{diam}$ &
$\rho$ & $k_{\text{max}}$\\
\midrule
{\emph{ LiveJournal      }  }    & 4,847,571        &68,993,773 & 16 & $\sim$ & $\sim$ \\
{\emph{ LiveJournal-Sym  }  }    & 4,847,571        &85,702,474 & 20 & 3480 & 372\\
{\emph{ com-Orkut        }  }    & 3,072,627        &234,370,166  & 9 & 5,667 & 253 \\
{\emph{ Twitter          }  }    & 41,652,231       &1,468,365,182 & 65* & $\sim$ & $\sim$ \\
{\emph{ Twitter-Sym      }  }    & 41,652,231       &2,405,026,092 & 23* & 14,963 & 2488\\
{\emph{ 3D-Torus         }  }    & 1,000,000,000    &6,000,000,000 & 1500* & 1 & 6\\
{\emph{ ClueWeb          }  }    & 978,408,098      &42,574,107,469 & 821* & $\sim$ & $\sim$\\
{\emph{ ClueWeb-Sym      }  }    & 978,408,098      &74,744,358,622 & 132* & 106,819 & 4244 \\
{\emph{ Hyperlink2014    }  }    & 1,724,573,718    &64,422,807,961 & 793* & $\sim$ & $\sim$ \\
{\emph{ Hyperlink2014-Sym}  }    & 1,724,573,718    &124,141,874,032 & 207* & 58,711 & 4160 \\
{\emph{ Hyperlink2012    }  }    & 3,563,602,789    &128,736,914,167 & 5275* & $\sim$ & $\sim$\\
{\emph{ Hyperlink2012-Sym}  }    & 3,563,602,789    &225,840,663,232 & 331* & 130,728 & 10565 \\
\end{tabular}
\captionof{table}{\small Graph inputs, including vertices and
  edges. $\mathsf{diam}$ is the diameter of the graph. For undirected
  graphs, $\rho$ and $k_{\text{max}}$ are the number of peeling rounds, and the largest non-empty core (degeneracy). We mark $\mathsf{diam}$ values where we are unable to calculate the exact diameter with * and report the effective diameter observed during our experiments, which is a lower bound on the actual diameter.}
\label{table:sizes}
\end{table}

\myparagraph{Graph Data}
To show how our algorithms perform on graphs at different scales, we
selected a representative set of real-world graphs of varying sizes. Most of
the graphs are Web graphs and social networks---low diameter graphs that are
frequently used in practice. To test our algorithms on large diameter
graphs, we also ran our implementations 3-dimensional tori where each
vertex is connected to its 2 neighbors in each dimension.

We list the graphs used in our experiments, along with their size,
approximate diameter, peeling complexity~\cite{dhulipala2017julienne},
and degeneracy (for undirected graphs) in
Table~\ref{table:sizes}. \defn{LiveJournal} is a directed graph of the
social network obtained from a snapshot in
2008~\cite{boldi2004webgraph}.  \defn{com-Orkut} is an undirected
graph of the Orkut social network.  \defn{Twitter} is a directed graph
of the Twitter network, where edges represent the follower
relationship~\cite{kwak2010twitter}.  \defn{ClueWeb} is a Web graph
from the Lemur project at
CMU~\cite{boldi2004webgraph}. \defn{Hyperlink2012} and
\defn{Hyperlink2014} are directed hyperlink graphs obtained from the
WebDataCommons dataset where nodes represent web
pages~\cite{meusel15hyperlink}. \defn{3D-Torus} is a 3-dimensional
torus with 1B vertices and 6B edges. We mark symmetric
(undirected) versions of the directed graphs with the
suffix -Sym.
We create weighted
graphs for evaluating weighted BFS, \Boruvka{}, widest path, and
Bellman-Ford by selecting edge weights between $[1, \log n)$ uniformly
at random. We process LiveJournal, com-Orkut, Twitter, and 3D-Torus in
the uncompressed format, and ClueWeb, Hyperlink2014, and Hyperlink2012
in the compressed format.

\begin{table*}[!t]
\footnotesize
\centering
\tabcolsep=0.12cm
\hspace*{-1em}
\begin{tabular}[t]{l | c|c|c | c|c|c | c|c|c | c|c|c }
  \toprule
  {\bf Application} &  \multicolumn{3}{c|}{LiveJournal-Sym} & \multicolumn{3}{c|}{com-Orkut} & \multicolumn{3}{c|}{Twitter-Sym}  & \multicolumn{3}{c}{3D-Torus}  \\
  & (1) & (72h) & (SU) & (1) & (72h) & (SU) & (1) & (72h) & (SU) & (1) & (72h) & (SU) \\
  \midrule
  {Breadth-First Search (BFS)}                 &0.59 &0.018 &32.7   &0.41  &0.012 &34.1   &5.45 &0.137 &39.7   &301  &5.53 &54.4     \\
  {Integral-Weight SSSP (weighted BFS)}        &1.45 &0.107 &13.5   &2.03  &0.095 &21.3   &33.4 &0.995 &33.5   &437  &18.1 &24.1     \\
  {General-Weight SSSP (Bellman-Ford)}         &3.39 &0.086 &39.4   &3.98  &0.168 &23.6   &48.7 &1.56  &31.2   &6280 &133  &47.2     \\
  {Single-Source Widest Path (Bellman-Ford)}                                &3.48 &0.090 &38.6   &4.39  &0.098 &44.7   &42.4 &0.749 &56.6   &580  &9.7  &59.7     \\
  {Single-Source Betweenness Centrality (BC)}  &1.66 &0.049 &33.8   &2.52  &0.057 &44.2   &26.3 &0.937 &28.0   &496  &12.5 &39.6     \\
  {$O(k)$-Spanner}                             &1.31 &0.041 &31.9   &2.34  &0.046 &50.8   &41.5 &0.768 &54.0   &380  &11.7 &32.4     \\
  {Low-Diameter Decomposition (LDD)}           &0.54 &0.027 &20.0   &0.33  &0.019 &17.3   &8.48 &0.186 &45.5   &275  &7.55 &36.4     \\
  {Connectivity}                               &1.01 &0.029 &34.8   &1.36  &0.031 &43.8   &34.6 &0.585 &59.1   &300  &8.71 &34.4     \\
  {Spanning Forest}                            &1.11 &0.035 &31.7   &1.84  &0.047 &39.1   &43.2 &0.818 &52.8   &334  &10.1 &33.0\\
  {Biconnectivity}                             &5.36 &0.261 &20.5   &7.31  &0.292 &25.0   &146  &4.86  &30.0   &1610 &59.6 &27.0     \\
  {Strongly Connected Components (SCC)*}       &1.61 &0.116 &13.8   &$\sim$&$\sim$&$\sim$ &13.3 &0.495 &26.8   &$\sim$&$\sim$&$\sim$ \\
  {Minimum Spanning Forest (MSF)}              &3.64 &0.204 &17.8   &4.58  &0.227 &20.1   &61.8 &3.02  &20.4   &617  &23.6 &26.1     \\
  {Maximal Independent Set (MIS)}              &1.18 &0.034 &34.7   &2.23  &0.052 &42.8   &34.4 &0.759 &45.3   &236  &4.44 &53.1     \\
  {Maximal Matching (MM)}                      &2.42 &0.095 &25.4   &4.65  &0.183 &25.4   &46.7 &1.42  &32.8   &403  &11.4 &35.3     \\
  {Graph Coloring}                             &4.69 &0.392 &11.9   &9.05  &0.789 &11.4   &148  &6.91  &21.4   &350  &11.3 &30.9     \\
  {Approximate Set Cover}                      &4.65 &0.613 &7.58   &4.51  &0.786 &5.73   &66.4 &3.31  &20.0   &1429 &40.2 &35.5     \\
  {$k$-core}                                   &3.75 &0.641 &5.85   &8.32  &1.33  &6.25   &110  &6.72  &16.3   &753  &6.58 &114.4    \\
  {Approximate Densest Subgraph}               &2.89 &0.052 &55.5   &4.71  &0.081 &58.1   &76.0 &1.14  &66.6   &95.4 &1.59 &60.0     \\
  {Triangle Counting (TC)}                     &13.5 &0.342 &39.4   &78.1  &1.19  &65.6   &1920 &23.5  &81.7   &168  &6.63 &25.3     \\
  {PageRank Iteration}                         &0.861 &0.012 &71.7  &1.28  &0.018 &71.1   &24.16 &0.453 &53.3  &107  &2.25 &47.5     \\
  \bottomrule
\end{tabular}
\caption{\small Running times (in seconds) of our algorithms over
  symmetric graph inputs on a 72-core machine (with hyper-threading)
  where (1) is the single-thread time, (72h) is the 72 core time using
  hyper-threading, and (SU) is the parallel speedup
  (single-thread time divided by 72-core time). We mark experiments
  that are not applicable for a graph with $\sim$, and experiments
  that did not finish within 5 hours with ---. *SCC was run on the directed versions of the input graphs. }
\label{table:small-times}
\end{table*}

\begin{table*}[!t]
\footnotesize
\centering
\tabcolsep=0.12cm
\hspace*{-0.2cm}
\begin{tabular}[t]{l | c|c|c | c|c|c | c|c|c}
  \toprule
  {\bf Application} &  \multicolumn{3}{c|}{ClueWeb-Sym} & \multicolumn{3}{c}{Hyperlink2014-Sym} & \multicolumn{3}{c}{Hyperlink2012-Sym} \\
  & (1) & (72h) & (SU) & (1) & (72h) & (SU) & (1) & (72h) & (SU)\\
  \midrule
  {Breadth-First Search (BFS)}                  &106  &2.29 &46.2       &250   &{4.50}  &55.5 	  & 576   & 8.44  & 68.2 \\
  {Integral-Weight SSSP (weighted BFS)}         &736  &14.4 &51.1       &1390  &{22.3} &62.3  	  & 3770  & 58.1  & 64.8 \\
  {General-Weight SSSP (Bellman-Ford)}          &1050 &16.2 &64.8       &1460  &{22.9} &63.7  	  & 4010  & 59.4  & 67.5 \\
  {Single-Source Widest Path (Bellman-Ford)}                                 &849  &11.8 &71.9       &1211  &16.8   &72.0      & 3210  & 48.4  & 66.3 \\
  {Single-Source Betweenness Centrality (BC)}   &569  &27.7 &20.5       &866   &{16.3} &53.1  	  & 2260  & 37.1  & 60.9 \\
  {$O(k)$-Spanner}                              &613  &9.79 &62.6       &906   &14.3   &63.3  	  & 2390  & 36.3  & 65.8 \\
  {Low-Diameter Decomposition (LDD)}            &176  &3.62 &48.6       &322   &{6.84} &47.0  	  & 980   & 16.6  & 59.0 \\
  {Connectivity}                                &381  &6.01 &63.3       &710   &{11.2} &63.3  	  & 1640  & 25.0  & 65.6 \\
  {Spanning Forest}                             &936  &18.2 &51.4       &1319  &{22.4} &58.8  	  & 2420  & 35.8  & 67.5 \\
  {Biconnectivity}                              &2250 &48.7 &46.2       &3520  &{71.5} &49.2  	  & 9860  & 165   & 59.7 \\
  {Strongly Connected Components (SCC)*}        &1240 &38.1 &32.5       &2140  &{51.5} &41.5  	  & 8130  & 185   & 43.9 \\
  {Minimum Spanning Forest (MSF)}               &2490 &45.6 &54.6       &3580  &{71.9} &49.7  	  & 9520  & 187   & 50.9 \\
  {Maximal Independent Set (MIS)}               &551  &8.44 &65.2       &1020  &{14.5} &70.3  	  & 2190  & 32.2  & 68.0 \\
  {Maximal Matching (MM)}                       &1760 &31.8 &55.3       &2980  &{48.1} &61.9  	  & 7150  & 108   & 66.2 \\
  {Graph Coloring}                              &2050 &49.8 &41.1       &3310  &{63.1} &52.4  	  & 8920  & 158   & 56.4 \\
  {Approximate Set Cover}                       &1490 &28.1 &53.0       &2040  &{37.6} &54.2  	  & 5320  & 90.4  & 58.8 \\
  {$k$-core}                                    &2370 &62.9 &37.6       &3480  &{83.2} &41.8  	  & 8515  & 184   & 46.0 \\
  {Approximate Densest Subgraph}                &1380 &19.6 &70.4       &1721  &24.3   &70.8  	  & 4420  & 61.4  & 71.9 \\
  {Triangle Counting (TC)}                      &13997 &204  &68.6      &---   &{480}  &---   	  & ---   & 1168  & --- \\
  {PageRank Iteration}                          &256.1 &3.49 &73.3      &385   &{5.17} &74.4  	  & 973   & 13.1 & 74.2 \\
  \bottomrule
\end{tabular}
\caption{\small Running times (in seconds) of our algorithms over
  symmetric graph inputs on a 72-core machine (with hyper-threading)
  where (1) is the single-thread time, (72h) is the 72 core time using
  hyper-threading, and (SU) is the parallel speedup
  (single-thread time divided by 72-core time). We mark experiments
  that are not applicable for a graph with $\sim$, and experiments
  that did not finish within 5 hours with ---. *SCC was run on the directed versions of the input graphs.}
\label{table:large-times}
\end{table*}

\subsection{SSSP Problems}
Our
BFS, weighted BFS, Bellman-Ford, and betweenness centrality
implementations achieve between a 13--67x speedup across
all inputs. We ran all of our shortest path experiments on the
\emph{symmetrized} versions of the graph. Our widest path
implementation achieves between 38--72x speedup across all inputs, and
our spanner implementation achieves between 31--65x speedup across all
inputs. We ran our spanner code with $k=4$. Our experiments show that our weighted BFS and Bellman-Ford
implementations perform as well as or better than our prior
implementations from Julienne~\cite{dhulipala2017julienne}. Our
running times for BFS and betweenness centrality are the same as the
times of the implementations in Ligra~\cite{shun2012ligra}. We note
that our running times for weighted BFS on the Hyperlink graphs are
larger than the times reported in Julienne. This is because the
shortest-path experiments in Julienne were run on directed version of
the graph, where the average vertex can reach many fewer vertices than
on the symmetrized version. We set a flag for our weighted BFS experiments on
the ClueWeb and Hyperlink graphs that lets the algorithm switch to a
dense \emap{} once the frontiers are sufficiently dense, which lets
the algorithm run within half of the RAM on our machine. Before this
change, our weighted BFS implementation would request a large amount of
amount of memory when processing the largest frontiers which then
caused the graph to become partly evicted from the page cache. For
widest path, the times we report are for the Bellman-Ford version of
the algorithm, which we were surprised to find is consistently
1.1--1.3x faster than our algorithm based on bucketing.
We observe that our spanner algorithm is only slightly more costly
than computing connectivity on the same input.

In an earlier paper~\cite{dhulipala2017julienne}, we
compared the running time of our weighted BFS implementation to two
existing parallel shortest path implementations from the GAP benchmark
suite~\cite{BeamerAP15} and Galois~\cite{maleki16dsmr}, as well as a
fast sequential shortest path algorithm from the DIMACS shortest path
challenge, showing that our implementation is between 1.07--1.1x slower
than the $\Delta$-stepping implementation from GAP, and 1.6--3.4x
faster than the Galois implementation. Our old version of Bellman-Ford
was between 1.2--3.9x slower than weighted BFS; we note that after
changing it to use the \emapblock{} optimization, it is now
competitive with weighted BFS and is between 1.2x faster and 1.7x
slower on our graphs with the exception of 3D-Torus, where it performs
7.3x slower than weighted BFS, as it performs $O(n^{4/3})$ work on this
graph.

\subsection{Connectivity Problems}
Our low-diameter decomposition (LDD) implementation achieves between
17--59x speedup across all inputs.
We fixed $\beta$ to $0.2$ in all of the codes that use LDD. The
running time of LDD is comparable to the cost of a BFS that visits
most of the vertices. We are not aware of any prior experimental work
that reports the running times for an LDD implementation.

Our work-efficient implementation of connectivity and spanning forest
achieve 25--57x speedup and 31--67x speedup across all inputs,
respectively.  We note that our implementation does not assume that
vertex IDs in the graph are randomly permuted and always generates a
random permutation, even on the first round, as adding vertices based
on their original IDs can result in poor performance (for example on
3D-Torus).  There are several existing implementations of fast
parallel connectivity algorithms~\cite{patwary2012cc, shun2012brief,
  shun2014practical, slota2014bfs}, however, only the implementation
from~\cite{shun2014practical}, which presents the connectivity
algorithm that we implement in this paper, is theoretically-efficient.
The implementation from Shun et al. was compared to both the
Multistep~\cite{slota2014bfs} and Patwary et al.~\cite{patwary2012cc}
implementations, and shown to be competitive on a broad set of graphs.
We compared our connectivity implementation to the work-efficient
connectivity implementation from Shun et al. on our uncompressed
graphs and observed that our code is between 1.2--2.1x faster in
parallel. Our spanning forest implementation is slightly slower than
connectivity due to having to maintain a mapping between the current
edge set and the original edge set.

Despite our biconnectivity implementation having $O(\mathsf{diam}(G))$ depth, our
implementation achieves between a 20--59x speedup across all inputs, as
the diameter of most of our graphs is extremely low. Our biconnectivity implementation is
about 3--5 times slower than running connectivity on the graph, which seems reasonable
as our current implementation performs two calls to connectivity, and one
breadth-first search. There are a several existing implementations of biconnectivity. Cong and Bader~\cite{cong2005experimental}
parallelize the Tarjan-Vishkin algorithm and demonstrated speedup over the
Hopcroft-Tarjan (HT) algorithm. Edwards and Vishkin~\cite{edwards2012bicc} also implement the
Tarjan-Vishkin algorithm using the XMT platform, and show that their algorithm
achieves good speedups. Slota and Madduri~\cite{slota2014simple} present a BFS-based biconnectivity
implementation which requires $O(mn)$ work in the worst-case, but behaves like
a linear-work algorithm in practice. We ran the Slota and Madduri implementation on 36 hyper-threads
allocated from the same socket, the configuration on which we observed the best
performance for their code, and found that our implementation is
between 1.4--2.1x faster than theirs. We used a DFS-ordered
subgraph corresponding to the largest connected component to test their code,
which produced the fastest times. Using the original order of the graph affects
the running time of their implementation, causing it to run between 2--3x slower
as the amount of work performed by their algorithm depends on the order in
which vertices are visited.

Our strongly connected components implementation achieves between a
13--43x
speedup across all inputs. Our implementation
takes a parameter $\beta$, which is the base of the exponential rate at which
we grow the number of centers added. We set $\beta$ between $1.1$--$2.0$ for our
experiments and note that using a larger value of $\beta$ can improve the running time on smaller graphs by up to a factor of 2x.
Our SCC implementation is between 1.6x faster to 4.8x slower than running connectivity on the
graph. There are several
existing SCC implementations that have been evaluated on real-world directed
graphs~\cite{hong2013scc, slota2014bfs, mclendon2005finding}. The Hong et al.
algorithm~\cite{hong2013scc} is a modified version of the FWBW-Trim algorithm from McLendon et
al.~\cite{mclendon2005finding}, but neither algorithm has any theoretical
bounds on work or depth. Unfortunately~\cite{hong2013scc} do not report
running times, so we are unable to compare our performance with them. The
Multistep algorithm~\cite{slota2014bfs} has a worst-case running time of $O(n^2)$,
but the authors point-out that the algorithm behaves like a linear-time
algorithm on real-world graphs. We ran our implementation on 16 cores configured
similarly to their experiments and found that we are about 1.7x slower on
LiveJournal, which easily fits in cache, and 1.2x faster on Twitter
(scaled to account for a small difference in graph sizes).  While the
multistep algorithm is slightly faster on some graphs, our SCC
implementation has the advantage of being theoretically-efficient and
performs a predictable amount of work.

Our minimum spanning forest implementation achieves between 17--54x speedup over
the implementation running on a single thread across all of our inputs.
Obtaining practical parallel algorithms for MSF has been a
longstanding goal in the field, and several existing implementations
exist~\cite{bader2006fast, nobari2012scalable, cong2016mst,
  shun2012brief,zhou2017mst}. We compared our
implementation with the union-find based MSF implementation from
PBBS~\cite{shun2012brief} and the implementation of \Boruvka{}
from~\cite{zhou2017mst}, which is one of the fastest
implementations we are aware of. Our MSF implementation is between
2.6--5.9x faster than the MSF implementation from PBBS. Compared to the
edgelist based implementation of \Boruvka{} from~\cite{zhou2017mst}
our implementation is between 1.2--2.9x faster.

\subsection{Covering Problems}
Our MIS and maximal matching implementations achieve between 31--70x
and 25--70x speedup across all inputs. The implementations by Blelloch
et al.~\cite{blelloch2012greedy} are the fastest existing
implementations of MIS and maximal matching that we are aware of, and
are the basis for our maximal matching implementation. They report
that their implementations are 3--8x faster than Luby's algorithm on
32 threads, and outperform a sequential greedy MIS implementation on
more than 2 processors. We compared our rootset-based MIS
implementation to the prefix-based implementation, and found that
the rootset-based approach is between 1.1--3.5x faster.
Our maximal matching implementation is between 3--4.2x faster than the
implementation from~\cite{blelloch2012greedy}. Our
implementation of maximal matching can avoid a significant amount of
work, as each of the filter steps can extract and permute just the
$3n/2$ highest priority edges, whereas the edgelist-based version in
PBBS must permute all edges. Our coloring implementation achieves
between 11--56x speedup across all inputs.  We note that our
implementation appears to be between 1.2--1.6x slower than the
asynchronous implementation of JP in~\cite{hasenplaugh2014ordering},
due to synchronizing on many rounds which contain few vertices.

Our approximate set cover implementation achieves between 5--57x
speedup across all inputs. Our implementation is based on the
implementation presented in Julienne~\cite{dhulipala2017julienne}; the
one major modification was to regenerate random priorities for sets that are
active on the current round. We compared the running time of our
implementation with the parallel implementation
from~\cite{blelloch12setcover} which is available in the PBBS
library. We ran both implementations with $\epsilon=0.01$. Our
implementation is between 1.2x slower to 1.5x faster than the PBBS
implementation on our graphs, with the exception of 3D-Torus. On
3D-Torus, the implementation from~\cite{blelloch12setcover} runs 56x
slower than our implementation as it does not regenerate priorities
for active sets on each round causing worst-case behavior. Our
performance is also slow on this graph, as nearly all of the vertices stay
active (in the highest bucket) during each round, and using
$\epsilon=0.01$ causes a large number of rounds to be performed.

\subsection{Substructure Problems}
Our $k$-core implementation achieves between 5--46x speedup across all inputs,
and 114x speedup on the 3D-Torus graph as there is only one round of
peeling in which all vertices are removed. There are several recent
papers that implement parallel algorithms for $k$-core~\cite{dasari2014park, dhulipala2017julienne, sariyuce2017parallel, Kabir2017}.
Both the ParK algorithm~\cite{dasari2014park}
and Kabir and Madduri algorithm~\cite{Kabir2017} implement the peeling
algorithm in $O(k_{\text{max}}n + m)$ work,
which is not work-efficient. Our implementation is between 3.8--4.6x faster than ParK on
a similar machine configuration. Kabir and Madduri show that their implementation
achieves an average speedup of 2.8x over ParK. Our implementation is between
1.3--1.6x faster than theirs on a similar machine configuration.

Our approximate densest subgraph implementation achieves between
44--77x speedup across all inputs. We ran our implementation with
$\epsilon = 0.001$, which in our experiments produced subgraphs with
density roughly equal to those produced by the 2-approximation
algorithm based on degeneracy ordering, or setting $\epsilon$ to 0. To
the best of our knowledge, there are no prior existing shared-memory
parallel algorithms for this problem.

Our triangle counting (TC) implementation achieves between 39--81x speedup across all
inputs. Unfortunately, we are unable to report speedup numbers for TC
on our larger graphs as the single-threaded times took too
long due to the algorithm performing $O(m^{3/2})$ work. There are a
number of experimental papers that consider multicore triangle
counting~\cite{Sevenich2014,Green2014,Kim2014,shun2015multicore,aberger2017empty,low2010graphlab}.
We implement the algorithm from~\cite{shun2015multicore}, and adapted
it to work on compressed graphs. We note that in our experiments we
intersect directed adjacency lists sequentially, as there was
sufficient parallelism in the outer parallel-loop. There was no
significant difference in running times between our implementation and
the implementation from~\cite{shun2015multicore}.  We ran our
implementation on 48 threads on the Twitter graph to compare with the
times reported by EmptyHeaded~\cite{aberger2017empty} and found that
our times are about the same.

\subsection{Eigenvector Problems}
Our PageRank (PR) implementation achieves between 39--54x speedup
across all inputs. Our implementation is based on the PageRank-Delta
implementation from Ligra~\cite{shun2012ligra}. We ran the algorithm
with $\epsilon = 1e-6$, and with $\epsilon' = 0.01$, where
$\epsilon'$ is the local (per-node) threshold for the amount of change
in its PageRank value before it is activated. We note that the
modification made to carry out dense iterations using a reduction over
the in-neighbors of a vertex was important to decrease contention, and
provided between 2--3x speedup over the Ligra implementation in
practice. Many graph processing systems implement PageRank. The
optimizing compiler used by GraphIt generates a highly-optimized implementation that is
currently the fastest shared-memory implementation known to us~\cite{graphit:2018}.
We note that our implementation is about 1.8x slower than the
implementation in GraphIt for LiveJournal and Twitter when run on the
same number of threads as in their experiments.

\begin{figure}[!t]
    \includegraphics[width=0.75\textwidth]{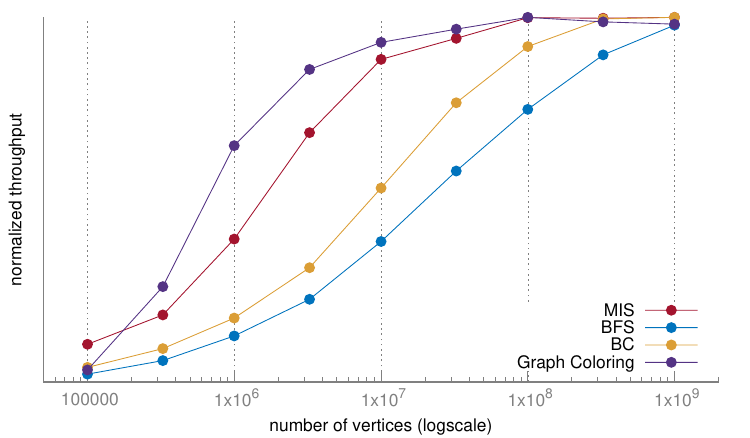}\\ 
    \caption{\small Log-linear plot of normalized throughput vs. vertices for MIS, BFS, BC, and coloring on the 3D-Torus graph family.}\label{fig:throughput}
\end{figure}

\subsection{Performance on 3D-Torus}
We ran experiments on a family of 3D-Torus graphs with different sizes
to study how our diameter-bounded algorithms scale relative to
algorithms with polylogarithmic depth. We were surprised to see that
the running time of some of our polylogarithmic depth algorithms on
this graph, like LDD and connectivity, are 17--40x more expensive than
their running time on Twitter and Twitter-Sym, despite 3D-Torus only
having 4x and 2.4x more edges than Twitter and Twitter-Sym. Our
slightly worse scaling on this graph can be accounted for by the fact
that we stored the graph ordered by dimension, instead of storing it
using a local ordering. It would be interesting to see how much
improvement we could gain by reordering the vertices.

In Figure~\ref{fig:throughput} we show the normalized throughput of
MIS, BFS, BC, and graph coloring for 3-dimensional tori of different
sizes, where throughput is measured as the number of edges
processed per second. The throughput for each application
becomes saturated before our largest-scale graph for all applications
except for BFS, which is saturated on a graph with 2 billion
vertices. The throughput curves show that the theoretical bounds are
useful in predicting how the half-lengths\footnote{The graph size when the
  system achieves half of its peak-performance.} are distributed.  The
half-lengths are ordered as follows: coloring, MIS, BFS, and BC. This is the
same order as sorting these algorithms by their depth with respect to this
graph.

\myparagraph{Locality} While our algorithms are efficient on the
\mpram{}, we do not analyze their cache complexity, and in general they may not be
efficient in a model that takes caches into account. Despite this, we
observed that our algorithms have good cache performance on the graphs
we tested on. In this section we give some explanation for this fact
by showing that our primitives make good use of the caches. Our
algorithms are also aided by the fact that these graph datasets often
come in highly local orders (e.g., see the \emph{Natural} order in
~\cite{dhulipala16compressing}).  Table~\ref{table:locality} shows
metrics for our experiments measured using Open Performance Counter
Monitor (PCM).

\setlength{\tabcolsep}{2pt}
\begin{table}[!t]\footnotesize
\centering
\centering
\begin{tabular}[!t]{l|c|c|c|c|c}
\toprule
Algorithm & Cycles Stalled & LLC Hit Rate & LLC Misses & BW & Time\\
\midrule
$k$-core (histogram)          & 9   & 0.223 & 49 & 96  & 62.9 \\
$k$-core (fetch-and-add)      & 67  & 0.155 & 42 & 24  & 221 \\
weighted BFS (blocked)        & 3.7 & 0.070 & 19 & 130 & 14.4 \\
weighted BFS (unblocked)      & 5.6 & 0.047 & 29 & 152 & 25.2 \\
\bottomrule
\end{tabular}
\captionof{table}{\small Cycles stalled while the memory subsystem has an
  outstanding load (trillions), LLC hit rate and misses (billions),
  bandwidth in GB/s (bytes read and written from memory, divided by
  running time), and running time in seconds. All experiments are run
  on the ClueWeb graph using 72 cores with hyper-threading.}
\label{table:locality}
\end{table}

Due to space limitations, we only report numbers for the
ClueWeb graph. We observed that using a work-efficient histogram is
3.5x faster than using fetch-and-add in our $k$-core implementation,
which suffers from high contention on this graph. Using a histogram
reduces the number of cycles stalled due to memory by more than 7x. We
also ran our wBFS implementation with and without the \emapblock{}
optimization, which reduces the number of cache-lines read from and
written to when performing a sparse \emap{}. The blocked
implementation reads and writes 2.1x fewer bytes than the unoptimized
version, which translates to a 1.7x faster runtime. We disabled the
dense optimization for this experiment to directly compare the two
implementations of a sparse \emap{}.

\subsection{Processing Massive Web Graphs}\label{subsec:massivecomparisons}
In Tables~\ref{table:small-times} and ~\ref{table:large-times}, we show the
running times of our implementations on the ClueWeb, Hyperlink2014, and
Hyperlink2012 graphs. To put our performance in context, we
compare our 72-core running times to running times reported
by existing work. Table~\ref{table:big-comparison} summarizes
state-of-the-art existing results in the literature. Most results
process the \emph{directed} versions of these graphs, which have about
half as many edges as the symmetrized version. Unless otherwise
mentioned, all results from the literature use the directed versions
of these graphs. To make the comparison easier we show our running
times for BFS, SSSP (weighted BFS), BC and SCC on the directed graphs,
and running times for Connectivity, $k$-core and TC on the symmetrized
graphs in Table~\ref{table:big-comparison}.

\setlength{\tabcolsep}{2pt}
\begin{table}[!t]\footnotesize
\centering

\begin{tabular}[!t]{lllrrrr}
\toprule
Paper & Problem & Graph & Memory & Hyper-threads & Nodes & Time \\
\midrule
\multirow{3}{*}{Mosaic~\cite{maass2017mosaic}}
& BFS*             & 2014 & 0.768 & 1000 & 1 & 6.55 \\ 
& Connectivity*    & 2014 & 0.768 & 1000 & 1 & 708  \\ 
& SSSP*            & 2014 & 0.768 & 1000 & 1 & 8.6  \\ 
\midrule

\multirow{4}{*}{FlashGraph~\cite{da2015flashgraph}}
& BFS*          & 2012 & .512 & 64 & 1 & 208  \\ 
& BC*           & 2012 & .512 & 64 & 1 & 595  \\ 
& Connectivity* & 2012 & .512 & 64 & 1 & 461  \\ 
& TC*           & 2012 & .512 & 64 & 1 & 7818 \\ 
\midrule

\multirow{2}{*}{BigSparse~\cite{jun2017bigsparse}}
& BFS* & 2012 & 0.064 & 32 & 1 & 2500  \\ 
& BC*  & 2012 & 0.064 & 32 & 1 & 3100  \\ 
\midrule

\multirow{3}{*}{Slota et al.~\cite{Slota2016}}
& Largest-CC*        & 2012 & 16.3 & 8192 & 256 & 63  \\    
& Largest-SCC*       & 2012 & 16.3 & 8192 & 256 & 108 \\   
& Approx $k$-core*   & 2012 & 16.3 & 8192 & 256 & 363 \\ 
\midrule

\multirow{1}{*}{Stergiou et al.~\cite{Stergiou2018}}
& Connectivity  & 2012 & 128 & 24000 & 1000 & 341 \\ 
\midrule

\multirow{1}{*}{Gluon~\cite{dathathri2018gluon}}
& BFS           & 2012 & 24 & 69632 & 256 & 380   \\
& Connectivity  & 2012 & 24 & 69632 & 256 & 75.3  \\
& PageRank      & 2012 & 24 & 69632 & 256 & 158.2 \\
& SSSP          & 2012 & 24 & 69632 & 256 & 574.9 \\
\midrule

\multirow{8}{*}{This paper}
& BFS*            & 2014 & 1 & 144 & 1 & 5.71 \\ 
& SSSP*           & 2014 & 1 & 144 & 1 & 9.08 \\ 
& Connectivity    & 2014 & 1 & 144 & 1 & 11.2 \\ 
& BFS*            & 2012 & 1 & 144 & 1 & 16.7 \\ 
& BC*             & 2012 & 1 & 144 & 1 & 35.2 \\ 
& Connectivity    & 2012 & 1 & 144 & 1 & 25.0 \\ 
& SCC*            & 2012 & 1 & 144 & 1 & 185  \\ 
& SSSP            & 2012 & 1 & 144 & 1 & 58.1 \\ 
& $k$-core        & 2012 & 1 & 144 & 1 & 184  \\ 
& PageRank        & 2012 & 1 & 144 & 1 & 462  \\ 
& TC              & 2012 & 1 & 144 & 1 & 1168 \\ 
\bottomrule

\end{tabular}
\captionof{table}{\small
  System configurations (memory in terabytes, hyper-threads, and nodes) and
  running times (seconds) of existing results on the Hyperlink graphs.
  The last section shows our running times. *These problems are run
  on directed versions of the graph.
} \label{table:big-comparison}
\end{table}

FlashGraph~\cite{da2015flashgraph} reports disk-based running times
for the Hyperlink2012 graph on a 4-socket,
32-core machine with 512GB of memory and 15 SSDs. On 64 hyper-threads, they
solve BFS in
208s, BC in 595s, connected components in 461s, and
triangle counting in 7818s. Our BFS
and BC implementations are 12x faster and 16x faster, and our triangle
counting and connectivity implementations are 5.3x faster
and 18x faster than their implementations, respectively.
Mosaic~\cite{maass2017mosaic}
report in-memory running times on the Hyperlink2014 graph; we note
that the system is optimized for external memory execution. They solve BFS in 6.5s, connected components
in 700s, and SSSP (Bellman-Ford) in 8.6s on a machine with 24
hyper-threads and 4 Xeon-Phis (244 cores with 4 threads each) for a
total of 1000 hyper-threads, 768GB of RAM,
and 6 NVMes. Our BFS and connectivity implementations are 1.1x and 62x
faster respectively, and our SSSP implementation is 1.05x slower. Both
FlashGraph and Mosaic compute weakly connected components, which is
equivalent to connectivity.
BigSparse~\cite{jun2017bigsparse} report disk-based running times for
BFS and BC on the Hyperlink2012 graph on a 32-core machine.  They
solve BFS in 2500s and BC in 3100s. Our BFS and BC implementations are
149x and 88x faster than their implementations, respectively.

Slota et al.~\cite{Slota2016} report running times for the Hyperlink2012 graph on
256 nodes on the Blue Waters supercomputer. Each node contains
two 16-core processors with one thread each, for a total of 8192
hyper-threads. They report they
can find the \emph{largest} connected component and SCC from the graph
in 63s and 108s respectively. Our implementations find \emph{all} connected
components 2.5x faster than their largest connected component
implementation, and find \emph{all}
strongly connected components 1.6x slower than their largest-SCC
implementation. Their largest-SCC implementation computes two BFSs
from a randomly chosen vertex---one on the in-edges and the other on the
out-edges---and intersects the reachable sets. We
perform the same operation as one of the first steps of our SCC
algorithm and note that it requires about 30 seconds on our machine. They solve
approximate $k$-cores in 363s, where the approximate $k$-core of a
vertex is the coreness of the vertex rounded up to the nearest powers
of 2. Our implementation computes the \emph{exact} coreness of each
vertex in 184s, which is 1.9x faster than the approximate
implementation while using 113x fewer cores.

Recently, Dathathri et al.~\cite{dathathri2018gluon} have reported
running times for the Hyperlink2012 graph using Gluon, a distributed
graph processing system based on Galois. They process this graph on a
256 node system, where each node is equipped with 68 4-way
hyper-threaded cores, and the hosts are connected by an Intel
Omni-Path network with 100Gbps peak bandwidth. They report times for
BFS, connectivity, PageRank, and SSSP. Other than their connectivity
implementation, which uses pointer-jumping, their implementations are
based on data-driven asynchronous label-propagation. We are not aware
of any theoretical bounds on the work and depth of these
implementations.  Compared to their reported times, our implementation
of BFS is 22.7x faster, our implementation of connectivity is 3x
faster, and our implementation of SSSP is 9.8x faster. Our PageRank
implementation is 2.9x slower. However, we note that the PageRank
numbers they report are not for true PageRank, but PageRank-Delta, and
are thus in some sense incomparable.

Stergiou et al.~\cite{Stergiou2018} describe a connectivity
algorithm that runs in $O(\log n)$ rounds in the BSP model and report
running times for the symmetrized Hyperlink2012 graph. They implement their
algorithm using a proprietary in-memory/secondary-storage graph
processing system used at Yahoo!, and run experiments on a
1000 node cluster. Each node contains two 6-core processors that are
2-way hyper-threaded and 128GB
of RAM, for a total of 24000 hyper-threads and 128TB of RAM. Their fastest
running time on the Hyperlink2012 graph is 341s on their 1000 node
system. Our implementation solves connectivity on this
graph in 25s--13.6x faster on a system with 128x less memory and 166x
fewer cores. They also report running times for solving connectivity
on a private Yahoo! webgraph with 272 billion vertices and 5.9 trillion
edges, over 26 times the size of our largest graph. While such a graph
seems to currently be out of reach of our machine, we are hopeful that
techniques from theoretically-efficient parallel algorithms can help
solve problems on graphs at this scale and beyond.

\section{Conclusion}
In this paper, we showed that we can process the largest
publicly-available real-world graph on a single shared-memory
server with 1TB of memory using theoretically-efficient parallel
algorithms. We outperform existing implementations on the largest
real-world graphs, and use much fewer resources than the
distributed-memory solutions. On a per-core basis, our numbers are
significantly better.  Our results provide evidence that
theoretically-efficient shared-memory graph algorithms can be
efficient and scalable in practice.

\section*{Acknowledgements}
Thanks to the reviewers and to Lin Ma for helpful comments.
This research was supported in part by
NSF grants \#CCF-1408940, \#CCF-1533858, \#CCF-1629444,  and \#CCF-1845763, and DOE grant
\#DE-SC0018947.

\bibliographystyle{abbrv}
\bibliography{ref}

\balance
\begin{appendix}
  \section{Graph Statistics}\label{sec:graphstats}

In this section, we list graph statistics computed for the graphs from
Section~\ref{sec:exps}.\footnote{Similar statistics can be
  found on the SNAP website (\url{https://snap.stanford.edu/data/})
  and the Laboratory for Web Algorithmics website
  (\url{http://law.di.unimi.it/datasets.php}).}
  These statistics include the number of connected components,
strongly connected components, colors used by the LLF and LF
heuristics, number of triangles, and several others. These
numbers will be useful for verifying the correctness or quality of
our algorithms in relation to future algorithms that also run on these graphs.
Although some of these numbers were present in
Table~\ref{table:sizes}, we include in the tables below for
completeness.

To ensure that the interested reader can reproduce these results we
provide details about the statistics that are not self-explanatory.
\begin{itemize}
  \item \emph{Effective Directed Diameter}: the maximum number of
    levels traversed during a graph traversal algorithm (BFS or SCC)
    on the unweighted directed graph.
  \item \emph{Effective Undirected Diameter}: the maximum number of
    levels traversed during a graph traversal algorithm (BFS) on the
    unweighted directed graph.
  \item \emph{Size of Largest
      (Connected/Biconnected/Strongly-Connected) Component}: The
    number of \emph{vertices} in the largest
    (connected/biconnected/strongly-connected) component. Note that in
    the case of biconnectivity, we assign labels to edges, so a vertex
    participates in a component for each distinct edge label incident
    to it.

  \item \emph{Num. Triangles}: The number of closed triangles in $G$,
    where each triangle $(u, v, w)$ is counted exactly once.

  \item \emph{Num. Colors Used by (LF/LLF)}: The number of colors used
    is just the maximum color ID assigned to any vertex.

  \item \emph{(Maximum Independent Set/Maximum Matching/Approximate Set Cover)
      Size}: We report the sizes of these objects computed by our
    implementations. For MIS and maximum matching we report this metric
    to lower-bound on the size of the maximum independent set and
    maximum matching supported by the graph. For approximate set
    cover, we run our code on instances similar to those used in prior
    work (e.g., Blelloch et al.~\cite{blelloch12setcover} and
    Dhulipala et al.~\cite{dhulipala2017julienne}) where the
    elements are vertices and the sets are the neighbors of each
    vertex in the undirected graph.  In the case of the social network
    and hyperlink graphs, this optimization problem naturally captures
    the minimum number of users or Web pages whose neighborhoods must
    be retrieved to cover the entire graph.

  \item \emph{$k_{\text{max}}$ (Degeneracy)}: The value of $k$ of the largest non-empty
    $k$-core.
\end{itemize}

\newpage

\begin{table*}[!t]
\footnotesize
\tabcolsep=0.25cm
\centering
\hspace*{-2em}
\begin{tabular}[t]{l  r}
  \toprule
  {\bf Statistic} &  Value\\
  \midrule
 	 {Num. Vertices}                                         &4,847,571 \\
 	 {Num. Directed Edges}                                   &68,993,773 \\
 	 {Num. Undirected Edges}                                 &85,702,474 \\
 	 {Effective Directed Diameter}                           &16 \\
 	 {Effective Undirected Diameter}                         &20 \\
   {Num. Connected Components}                             &1,876 \\
   {Num. Biconnected Components}                           &1,133,883 \\
   {Num. Strongly Connected Components}                    &971,232 \\
   {Size of Largest Connected Component}                   &4,843,953 \\
   {Size of Largest Biconnected Component}                 &3,665,291 \\
   {Size of Largest Strongly Connected Component}          &3,828,682 \\
   {Num. Triangles}                                        &285,730,264 \\
   {Num. Colors Used by LF}                           		 &323 \\
   {Num. Colors Used by LLF}                           		 &327 \\
   {Maximal Independent Set Size}                          &2,316,617 \\
   {Maximal Matching Size}                                 &1,546,833 \\
   {Set Cover Size}                                        &964,492 \\
   {$k_{\text{max}}$ (Degeneracy)}                         &372 \\
   {$\rho$ (Num. Peeling Rounds in $k$-core)}              &3,480 \\
  \bottomrule
\end{tabular}
\caption{\small Graph statistics for the LiveJournal graph.}
\label{table:livejournalstats}
\end{table*}

\begin{table*}[!t]
\footnotesize
\tabcolsep=0.25cm
\centering
\hspace*{-2em}
\begin{tabular}[t]{l  r}
  \toprule
  {\bf Statistic} &  Value\\
  \midrule
 	 {Num. Vertices}                                         &3,072,627 \\
 	 {Num. Directed Edges}                                   &--- \\
 	 {Num. Undirected Edges}                                 &234,370,166 \\
 	 {Effective Directed Diameter}                           &--- \\
 	 {Effective Undirected Diameter}                         &9 \\
   {Num. Connected Components}                             &187 \\
   {Num. Biconnected Components}                           &68,117 \\
   {Num. Strongly Connected Components}                    &--- \\
   {Size of Largest Connected Component}                   &3,072,441 \\
   {Size of Largest Biconnected Component}                 &3,003,914 \\
   {Size of Largest Strongly Connected Component}          &--- \\
   {Num. Triangles}                                        &627,584,181 \\
   {Num. Colors Used by LF}                           		 &86 \\
   {Num. Colors Used by LLF}                           		 &98 \\
   {Maximal Independent Set Size}                          &651,901 \\
   {Maximal Matching Size}                                 &1,325,427 \\
   {Set Cover Size}                                        &105,572 \\
   {$k_{\text{max}}$ (Degeneracy)}                         &253 \\
   {$\rho$ (Num. Peeling Rounds in $k$-core)}              &5,667 \\
  \bottomrule
\end{tabular}
\caption{\small Graph statistics for the com-Orkut graph. As com-Orkut
is an undirected graph, some of the statistics are not applicable and
we mark the corresponding values with --.}
\label{table:orkutstats}
\end{table*}

\begin{table*}[!t]
\footnotesize
\tabcolsep=0.25cm
\centering
\hspace*{-2em}
\begin{tabular}[t]{l  r}
  \toprule
  {\bf Statistic} &  Value\\
  \midrule
 	 {Num. Vertices}                                         &41,652,231 \\
 	 {Num. Directed Edges}                                   &1,468,365,182 \\
 	 {Num. Undirected Edges}                                 &2,405,026,092 \\
 	 {Effective Directed Diameter}                           &65 \\
 	 {Effective Undirected Diameter}                         &23 \\
   {Num. Connected Components}                             &2 \\
   {Num. Biconnected Components}                           &1,936,001 \\
   {Num. Strongly Connected Components}                    &8,044,729 \\
   {Size of Largest Connected Component}                   &41,652,230 \\
   {Size of Largest Biconnected Component}                 &39,708,003 \\
   {Size of Largest Strongly Connected Component}          &33,479,734 \\
   {Num. Triangles}                                        &34,824,916,864 \\
   {Num. Colors Used by LF}                           		 &1,081 \\
   {Num. Colors Used by LLF}                           		 &1,074 \\
   {Maximal Independent Set Size}                          &26,564,540 \\
   {Maximal Matching Size}                                 &9,612,260 \\
   {Set Cover Size}                                        &1,736,761 \\
   {$k_{\text{max}}$ (Degeneracy)}                         &2,488 \\
   {$\rho$ (Num. Peeling Rounds in $k$-core)}              &14,963 \\
  \bottomrule
\end{tabular}
\caption{\small Graph statistics for the Twitter graph.}
\label{table:twitterstats}
\end{table*}

\begin{table*}[!t]
\footnotesize
\tabcolsep=0.25cm
\centering
\hspace*{-2em}
\begin{tabular}[t]{l  r}
  \toprule
  {\bf Statistic} &  Value\\
  \midrule
 	 {Num. Vertices}                                         &978,408,098\\
 	 {Num. Directed Edges}                                   &42,574,107,469 \\
 	 {Num. Undirected Edges}                                 &74,774,358,622 \\
 	 {Effective Directed Diameter}                           &821 \\
 	 {Effective Undirected Diameter}                         &132 \\
   {Num. Connected Components}                             &23,794,336 \\
   {Num. Biconnected Components}                           &81,809,602 \\
   {Num. Strongly Connected Components}                    &135,223,661 \\
   {Size of Largest Connected Component}                   &950,577,812 \\
   {Size of Largest Biconnected Component}                 &846,117,956 \\
   {Size of Largest Strongly Connected Component}          &774,373,029 \\
   {Num. Triangles}                                        &1,995,295,290,765 \\
   {Num. Colors Used by LF}                           		 &4,245 \\
   {Num. Colors Used by LLF}                           		 &4,245 \\
   {Maximal Independent Set Size}                          &459,052,906 \\
   {Maximal Matching Size}                                 &311,153,771 \\
   {Set Cover Size}                                        &64,322,081 \\
   {$k_{\text{max}}$ (Degeneracy)}                         &4,244 \\
   {$\rho$ (Num. Peeling Rounds in $k$-core)}              &106,819 \\
  \bottomrule
\end{tabular}
\caption{\small Graph statistics for the ClueWeb graph.}
\label{table:cluewebstats}
\end{table*}

\begin{table*}[!t]
\footnotesize
\tabcolsep=0.25cm
\centering
\hspace*{-2em}
\begin{tabular}[t]{l  r}
  \toprule
  {\bf Statistic} &  Value\\
  \midrule
   {Num. Vertices}                                         &1,724,573,718\\
   {Num. Directed Edges}                                   &64,422,807,961 \\
   {Num. Undirected Edges}                                 &124,141,874,032 \\
   {Effective Directed Diameter}                           &793 \\
   {Effective Undirected Diameter}                         &207 \\
   {Num. Connected Components}                             &129,441,050 \\
   {Num. Biconnected Components}                           &132,198,693 \\
   {Num. Strongly Connected Components}                    &1,290,550,195 \\
   {Size of Largest Connected Component}                   &1,574,786,584 \\
   {Size of Largest Biconnected Component}                 &1,435,626,698 \\
   {Size of Largest Strongly Connected Component}          &320,754,363 \\
   {Num. Triangles}                                        &4,587,563,913,535 \\
   {Num. Colors Used by LF}                           		 &4154 \\
   {Num. Colors Used by LLF}                           		 &4158 \\
   {Maximal Independent Set Size}                          &1,333,026,057 \\
   {Maximal Matching Size}                                 &242,469,131 \\
   {Set Cover Size}                                        &23,869,788 \\
   {$k_{\text{max}}$ (Degeneracy)}                         &4,160 \\
   {$\rho$ (Num. Peeling Rounds in $k$-core)}              &58,711 \\
  \bottomrule
\end{tabular}
\caption{\small Graph statistics for the Hyperlink2014 graph.}
\label{table:hyperlink2014stats}
\end{table*}

\begin{table*}[!t]
\footnotesize
\tabcolsep=0.25cm
\centering
\hspace*{-2em}
\begin{tabular}[t]{l  r}
  \toprule
  {\bf Statistic} &  Value\\
  \midrule
   {Num. Vertices}                                         &3,563,602,789  \\
   {Num. Directed Edges}                                   &128,736,914,167 \\
   {Num. Undirected Edges}                                 &225,840,663,232 \\
 	 {Effective Directed Diameter}                           &5275 \\
 	 {Effective Undirected Diameter}                         &331 \\
   {Num. Connected Components}                             &144,628,744\\
   {Num. Biconnected Components}                           &298,663,966 \\
   {Num. Strongly Connected Components}                    &1,279,696,892 \\
   {Size of Largest Connected Component}                   &3,355,386,234\\
   {Size of Largest Biconnected Component}                 &3,023,064,231 \\
   {Size of Largest Strongly Connected Component}          &1,827,543,757 \\
   {Num. Triangles}                                        &9,648,842,110,027 \\
   {Num. Colors Used by LF}                           		 &10,566 \\
   {Num. Colors Used by LLF}                           		 &10,566 \\
   {Maximal Independent Set Size}                          &1,799,823,993\\
   {Maximal Matching Size}                                 &2,434,644,438\\
   {Set Cover Size}                                        &372,668,619\\
   {$k_{\text{max}}$ (Degeneracy)}                         &10,565\\
   {$\rho$ (Num. Peeling Rounds in $k$-core)}              &130,728\\
  \bottomrule
\end{tabular}
\caption{\small Graph statistics for the Hyperlink2012 graph.}
\label{table:hyperlink2012stats}
\end{table*}

\end{appendix}

\end{document}